\documentclass[showpacs,showkeys,preprintnumbers,amsmath,amssymb,superscriptaddress,nofootinbib]{revtex4-1}

\usepackage{graphicx}
\usepackage{dcolumn}
\usepackage{bm}
\usepackage{natbib}
\usepackage[usenames,dvipsnames]{xcolor}
\usepackage{pifont}
\usepackage{relsize}
\definecolor{lightgreen}{rgb}{0,1,0}
\definecolor{darkgray}{gray}{0.20}
\usepackage{subfigure}
\usepackage{hyperref}
\usepackage{epstopdf}
\usepackage{float} 
\usepackage{multirow} 
\usepackage{afterpage} 

\begin{document}

\title{World Input-Output Network}

\author{Federica Cerina} \affiliation{Department of Physics,
  Universit\`{a} degli Studi di Cagliari, Cagliari, Italy}\affiliation{Linkalab, Complex
  Systems Computational Laboratory, Cagliari 09129, Italy} 
\author{Zhen Zhu} \affiliation{IMT Institute 
  for Advanced Studies Lucca, Piazza S. Ponziano 6, 55100 Lucca, Italy} 
\author{Alessandro Chessa}  \affiliation{Linkalab, Complex
  Systems Computational Laboratory, Cagliari 09129, Italy} \affiliation{IMT Institute 
  for Advanced Studies Lucca, Piazza S. Ponziano 6, 55100 Lucca, Italy}     
\author{Massimo Riccaboni} \affiliation{IMT Institute 
  for Advanced Studies Lucca, Piazza S. Ponziano 6, 55100 Lucca, Italy} \affiliation{DMSI, KU Leuven, Belgium}

\begin{abstract}
\noindent Economic systems, traditionally analyzed as almost independent national systems, are increasingly connected on a global scale. Only recently becoming available, the World Input-Output Database (WIOD) is one of the first efforts to construct the multi-regional input-output (MRIO) tables at the global level. By viewing the world input-output system as an interdependent network where the nodes are the individual industries in different economies and the edges are the monetary goods flows between industries, we study the network properties of the so-called world input-output network (WION) and document its evolution over time. We are able to quantify not only some global network properties such as assortativity, clustering coefficient, and degree and strength distributions, but also its subgraph structure and dynamics by using community detection techniques. Over time, we detect a marked increase in cross-country connectivity of the production system, only temporarily interrupted by the 2008-2009 crisis. Moreover, we find a growing input-output regional community in Europe led by Germany and the rise of China in the global production system. Finally, we use the network-based PageRank centrality and community coreness measure to identify the key industries and economies in the WION and the results are different from the one obtained by the traditional final-demand-weighted backward linkage measure. 
\end{abstract}

\date{\today}



\maketitle

\section{Introduction} \label{sec:intro}

As the global economy becomes increasingly integrated, an isolated view based on the national input-output table\footnote{Ever since Leontief \cite{leontief1936quantitative} formalized its structure, the input-output table has been used extensively by economists, environmentalists, and policy makers alike. By keeping track of the inter-industrial relationships, the input-output table offers a reasonably accurate measurement of the response of any given economy in the face of external shocks or policy interventions.} is no longer sufficient to assess an individual economy's strength and weakness, not to mention finding solutions to global challenges such as climate change and financial crises. Hence, a multi-regional input-output (MRIO) framework is needed to draw a high-resolution representation of the global economy \cite{wiedmann2011quo}. In practice, however, due to the expensive process of collecting data and the variety of classifications used by different agencies, for a long time, the input-output tables have only been available for a limited number of countries and for discontinuous years. Fortunately, the fully-fledged MRIO databases started to become available in recent years\footnote{Tukker and Dietzenbacher \cite{tukker2013global} summarize the recent development of the MRIO databases.}.

Unlike the national input-output table where exports and imports are aggregated and appended to final demand and country-specific value added respectively, for each individual economy, the MRIO table splits its exports into intermediate use and final use in every foreign economy and also traces its imports back to the industry origins in every foreign economy. As a result, the inter-industrial relationships in the MRIO table are recorded not only within the same economy but also across economies.  

The availability of the MRIO databases was soon followed by a wave of empirical studies of topics ranging from global value chains and trade fragmentation in economics \cite{baldwin2013supply,timmer2013fragmentation,koopman2014tracing} to global environmental accounting in ecology and resources management \cite{lenzen2012international,lenzen2013international,wiedmann2013material}. However, to the best of our knowledge, our paper is the first attempt to explore the MRIO tables from a networks perspective, even though there have been some networks studies of the input-output tables at the national level and for selected countries \cite{blochl2011vertex,mcnerney2013network,contreras2014}.   

Complex networks are a modern way to characterize mathematically a series of different systems in the shape of subunits (nodes) connected by their interaction (edges) \cite{ABRMP,SIAM}. Such modeling has been proved to be fruitful for the description 
of a variety of different phenomena ranging from biology \cite{GCbook2} to economics \cite{kitsak2010,pammolli2002,riccaboni2010,riccaboni2013,chessa2013} and 
finance \cite{nature}. Here we move forward by considering the global MRIO system as a world input-output network (WION), where the nodes are the individual industries in different economies and the edges are the monetary goods flows\footnote{More precisely, the edges are the monetary goods and services flows. The direction of the flows go from the seller industry to the buyer industry. They are monetary because they are denoted in current US dollars.} between industries, similarly to what have been done recently by Acemoglu, Carvalho, Ozdaglar and Tahbaz-Salehi for the US economy only \cite{acemoglu2012network}. 

Different from many network systems observed in reality, the WION has the following features: 1) It is directed and weighted, i.e., an industry can act as both a seller and a buyer at the same time and the monetary goods flows between industries vary a lot; 2) It is much denser within the same economy than across economies, i.e., despite the continuously integrated global economy, most economic transactions still happen within the country border;\footnote{In contrast, due to the low-digit industry classification, the input-output networks at the national level are almost complete \cite{blochl2011vertex}.} 3) It is with strong self-loops, i.e., an industry can acquire a significant amount of inputs from itself\footnote{This is also due to the aggregated industry classification.}.

Taking into account the features above, we explore the WION by quantifying not only some global network properties such as assortativity but also some local network properties such as PageRank centrality. Furthermore, we apply community detection and community core detection techniques to examine the structure of the WION over time.

This paper makes some significant contributions to the literature of input-output economics. First, it is the first attempt to quantify the network properties of the WION by taking into account its edge weights and directedness\footnote{Carvalho \cite{carvalho2013survey} also use a networks approach to study the WIOD data. But he only uses a single year (2006) and considers it as an unweighted network.}. By doing that, we detect a marked increase in cross-country connectivity, apart from a sharp drop in 2009 due to the financial crisis. Second, the community detection results reveal growing input-output international communities. Among them, we notice in particular the emergence of a large European community led by Germany. Third, we use the network-based PageRank centrality and community coreness measure to identify the key industries and economies in the WION and the results are different from the one obtained by the traditional final-demand-weighted backward linkage measure.

The rest of the paper is structured as follows. Section \ref{sec:data} describes the database used and the MRIO framework. We also conduct a basic MRIO analysis to identify the key industries at the global level in this section. Section \ref{sec:properties} quantifies some global network properties of the WION and its subgraph structure and dynamics by using community detection techniques. Moreover, we use the network-based PageRank centrality and community coreness measure to identify the key industries in the WION.   
Finally, Section \ref{sec:conclusion} concludes the paper.

\section{The Data Description and the Leontief-Inverse-Based Method of Identifying the Key Industries} \label{sec:data}

\subsection{The WIOD Data and the MRIO Framework}

We use the World Input-Output Database (WIOD) \cite{timmer2012world} to map out the WION. At the time of writing, the WIOD input-output tables cover 35 industries for each of the 40 economies (27 EU countries and 13 major economies in other regions) plus the rest of the world (RoW) and the years from 1995 to 2011\footnote{Tables \ref{table_A1} and \ref{table_A2} in the appendix have the lists of countries and industries covered in the WIOD.}. For each year, there is a harmonized global level input-output table recording the input-output relationships between any pair of industries in any pair of economies\footnote{Again, the relationship can also be an industry to itself and within the same economy.}. The numbers in the WIOD are in current basic (producers') prices and are expressed in millions of US dollars\footnote{The basic prices are also called the producers' prices, which represent the amount receivable by the producers. An alternative is the purchases' prices, which represent the amount paid by the purchases and often include trade and transport margins. The former is preferred by the WIOD because it better reflects the cost structures underlying the industries \cite{timmer2012world}.}. Table \ref{table_1} shows an example of a MRIO table with two economies and two industries. The $4\times4$ inter-industry table is called the transactions matrix and is often denoted by $\boldsymbol{\mathrm{Z}}$. The rows of $\boldsymbol{\mathrm{Z}}$ record the distributions of the industry outputs throughout the two economies while the columns of $\boldsymbol{\mathrm{Z}}$ record the composition of inputs required by each industry. Notice that in this example all the industries buy inputs from themselves, which is often observed in real data. Besides intermediate industry use, the remaining outputs are absorbed by the additional columns of final demand, which includes household consumption, government expenditure, and so forth\footnote{Here we only show the aggregated final demand for the two economies}. Similarly, production necessitates not only inter-industry transactions but also labor, management, depreciation of capital, and taxes, which are summarized as the additional row of value added. The final demand matrix is often denoted by $\boldsymbol{\mathrm{F}}$ and the value added vector is often denoted by $\boldsymbol{\mathrm{v}}$. Finally, the last row and the last column record the total industry outputs and its vector is denoted by $\boldsymbol{\mathrm{x}}$. 

\afterpage{
\begin{table}
\caption{{\bf A hypothetical two-economy-two-industry MRIO table.} The $4\times4$ inter-industry transactions matrix records outputs selling in its rows and inputs buying in its columns. The additional columns are the final demand and the additional row is the value added. Finally, the last column and the last row record the total industry outputs. The numbers are made up in such a way that Economy 2 is a lot larger than Economy 1 in terms of industry outputs. However, as shown below, an unweighted backward linkage measure will consider industries in Economy 1 more important than the ones in Economy 2. Hence, we adopt a final-demand-weighted backward linkage measure to identify the key industries in the WIOD.}
  \centering
  \resizebox{18cm}{!}{
    \begin{tabular}{@{\extracolsep{4pt}}ccccccccc@{}}
    \toprule
          &       & \multicolumn{4}{c}{Buyer Industry} &       &       &  \\
        \cline{3-6}
          &       & \multicolumn{2}{c}{Economy 1} & \multicolumn{2}{c}{Economy 2} & \multicolumn{2}{c}{Final Demand} &  \\
          \cline{3-4} \cline{5-6} \cline{7-8}
          \multicolumn{2}{c}{Seller Industry}        & Industry 1 & Industry 2 & Industry 1 & Industry 2 & Economy 1 & Economy 2 & Total Output \\
          \colrule
     \multirow{2}{*}{Economy 1} & Industry 1 & 5    & 10     & 20    & 10    & 45     & 10     & 100 \\
      & Industry 2 & 10     & 5    & 10    & 20    & 50     & 5     & 100 \\
    \multirow{2}{*}{Economy 2} & Industry 1 & 30     & 15     & 800  & 500   & 5  & 8650  & 10000 \\
     & Industry 2 & 35     & 30     & 1000   & 1000   & 25  & 7910  & 10000 \\
      \multicolumn{2}{c}{Value Added}  & 20    & 40    & 8170  & 8470  &       &       &  \\
     \multicolumn{2}{c}{Total Output} & 100    & 100    & 10000 & 10000 &       &       &  \\
          \botrule
    \end{tabular}%
    }
  \label{table_1}%
\end{table}
\begin{figure}[!t]
\centering
{\includegraphics[width=7cm]{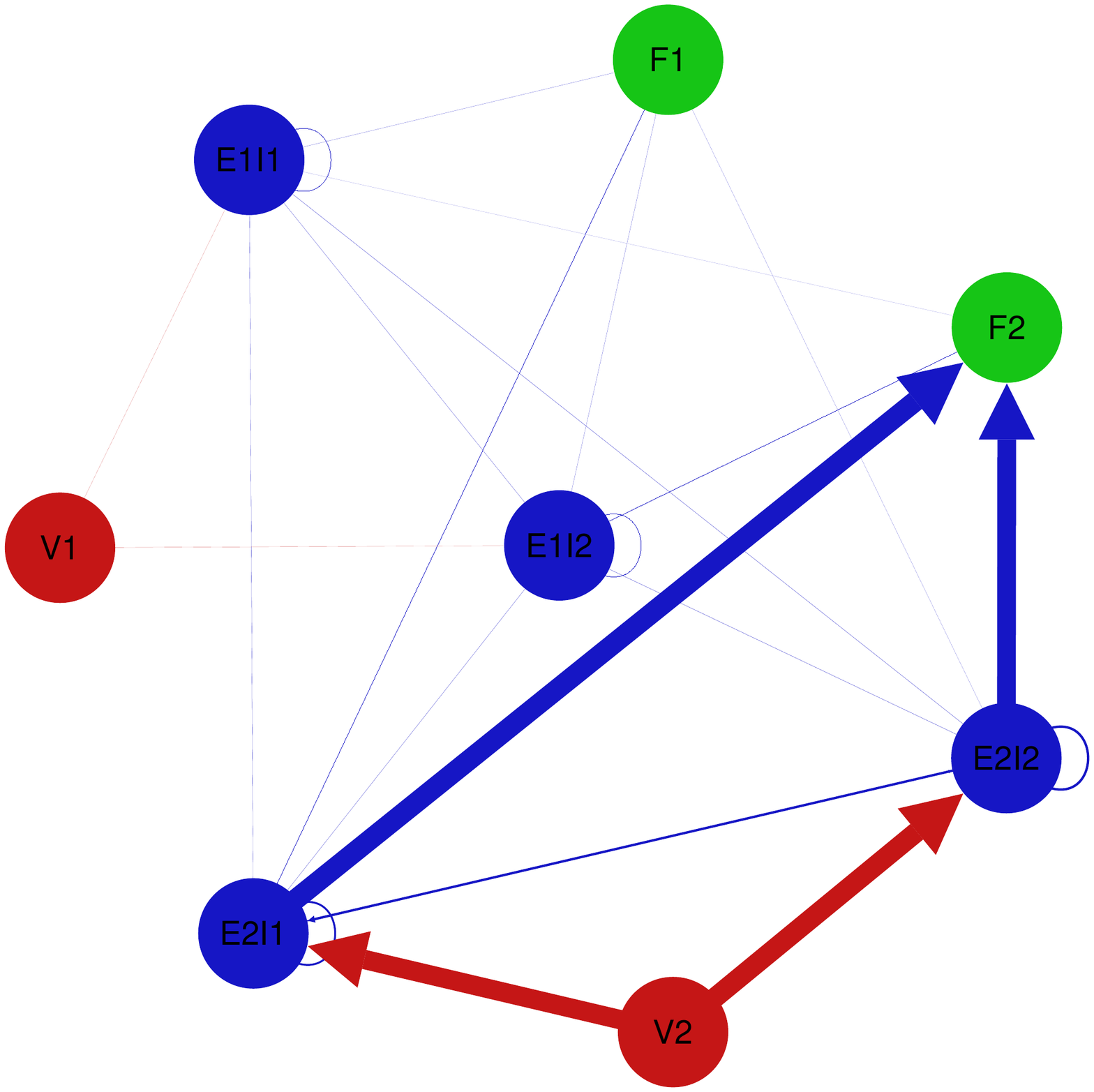}}
\caption{{\bf A hypothetical two-economy-two-industry WION.} This is a topological view of Table \ref{table_1}. The blue nodes are the individual industries. The label ``ExIy'' should read ``Industry y in Econoomy x''. The red nodes are the value added sources from the two economies, whereas the green nodes are the final demand destinations in the two economies. The label ``Vx'' should read ``Value Added from Economy x'', whereas the label ``Fx'' should read ``Final Demand in Economy x''. The edges are with arrows indicating the directions of the monetary goods flows and with varying widths indicating the magnitudes of the flows. The color of the edge is set the same as the source node's. Finally, because we are only concerned with the inter-industrial input-output relationships, when formulating the WION, we focus our attention on the network among the blue nodes.} \label{table1Net}
\end{figure}
}

\subsection{The Leontief-Inverse-Based Method of Identifying the Key Industries}

If we use $\boldsymbol{\mathrm{i}}$ to denote a summation vector of conformable size, i.e., a vector of all 1's with the length conformable to the multiplying matrix, and let $\boldsymbol{\mathrm{F}}\boldsymbol{\mathrm{i}}=\boldsymbol{\mathrm{f}}$, we then have $\boldsymbol{\mathrm{Z}}\boldsymbol{\mathrm{i}}+\boldsymbol{\mathrm{f}}=\boldsymbol{\mathrm{x}}$. Furthermore, if dividing each column of $\boldsymbol{\mathrm{Z}}$ by its corresponding total output in $\boldsymbol{\mathrm{x}}$, we get the so-called technical coefficients matrix $\boldsymbol{\mathrm{A}}$\footnote{The ratios are called technical coefficients because they represent the technologies employed by the industries to transform inputs into outputs.}. Replacing $\boldsymbol{\mathrm{Z}}\boldsymbol{\mathrm{i}}$ with $\boldsymbol{\mathrm{A}}\boldsymbol{\mathrm{x}}$, we rewrite the above equation as $\boldsymbol{\mathrm{A}}\boldsymbol{\mathrm{x}}+\boldsymbol{\mathrm{f}}=\boldsymbol{\mathrm{x}}$. It can be rearranged as $(\boldsymbol{\mathrm{I}}-\boldsymbol{\mathrm{A}})\boldsymbol{\mathrm{x}}=\boldsymbol{\mathrm{f}}$. Finally, we can solve $\boldsymbol{\mathrm{x}}$ as follows:
\begin{equation}
\boldsymbol{\mathrm{x}}=(\boldsymbol{\mathrm{I}}-\boldsymbol{\mathrm{A}})^{-1}\boldsymbol{\mathrm{f}}
\end{equation}
where matrix $(\boldsymbol{\mathrm{I}}-\boldsymbol{\mathrm{A}})^{-1}$ is often denoted by $\boldsymbol{\mathrm{L}}$ and is called the Leontief inverse. 

The intuition behind the Leontief inverse is that an increase in the final demand of an industry's output will induce not only more production from the industry itself but also more from other related industries because more inputs are required. Therefore, the Leontief inverse takes into account both the direct and indirect effects of a demand increase. For instance, $L_{ij}$ measures the total output produced in Industry $i$ given a one-unit increase in Industry $j$'s final demand\footnote{The Leontief inverse is demand-driven, i.e., a repercussion effect triggered by an increase in final demand. Another strand of the input-output economics literature is based on the supply-driven model, where a repercussion effect is triggered by an increase in value added (primary inputs) \cite{ghosh1958input,miller2009input}.}. As a result, $\boldsymbol{\mathrm{i}}'\boldsymbol{\mathrm{L}}$ sums up each column of $\boldsymbol{\mathrm{L}}$ and each sum measures the total output of all the industries given a one-unit increase in the corresponding industry's final demand. The vector $\boldsymbol{\mathrm{i}}'\boldsymbol{\mathrm{L}}$ is called the backward linkage measure\footnote{It is backward because the linkage is identified by tracing back to the upstream industries.} and can be used to rank the industries and identify the key ones in the economy \cite{yotopoulos1973balanced}. However, as pointed out by Laumas \cite{laumas1976weighting}, the key assumption embedded in the backward linkage measure is that every industry is assigned with the same weight (or unweighted), which is far from the reality. The problem with the unweighted backward linkage measure can be demonstrated by using the hypothetical data from Table \ref{table_1}. The calculated $\boldsymbol{\mathrm{i}}'\boldsymbol{\mathrm{L}}$ is $\begin{bmatrix}
2.0688 & 1.8377 & 1.2223 & 1.1854
\end{bmatrix}$, which considers the industries in Economy 1 more important than the ones in Economy 2, despite the fact that Economy 2 is a lot larger than Economy 1 in terms of total outputs. 

The industries of the 40 economies covered in the WIOD are very heterogeneous in terms of both total outputs and technical structure, which certainly makes the unweighted backward linkage measure not a good choice to identify the most central industries on a global scale. In order to identify the key industries in the WIOD, we hence follow Laumas \cite{laumas1976weighting} and use the final-demand-weighted backward linkage measure, which is denoted by $\boldsymbol{\mathrm{w}}$ and is defined here as the Hadamard (element-wise) product of the vector of the unweighted backward linkage measure and the vector of the percentage shares of the total final demand across industries, i.e.,
\begin{equation}
\boldsymbol{\mathrm{w}}=\boldsymbol{\mathrm{i}}'\boldsymbol{\mathrm{L}}\circ\frac{\boldsymbol{\mathrm{f}}'}{\boldsymbol{\mathrm{i}}'\boldsymbol{\mathrm{f}}}
\end{equation} 
where $\circ$ is the element-wise multiplication operator.  

Table \ref{table_2} shows the top 20 industries for the years 1995, 2003, and 2011, respectively. The first column of each year is produced by the final-demand-weighted backward linkage measure, i.e., $\boldsymbol{\mathrm{w}}$. For the selected years, only four large economies, China, Germany, Japan, and USA, ever qualified for the top 20. Another noticeable change over time is the rise of China, which topped the list in 2011 with its industry of construction. 

Tables \ref{table_A3} and \ref{table_A4} in the appendix provide an alternative way of viewing the key industries and economies over time identified by the final-demand-weighted backward linkage measure. In particular, Table \ref{table_A3} lists the most important economies by industry while Table \ref{table_A4} lists the most important industries by economy.

\section{The Network Properties of the WION and the Network-Based Methods of Identifying the Key Industries} \label{sec:properties}

As mentioned in Section \ref{sec:intro}, the complex networks approach has been widely used in economics and finance in recent years \cite{kitsak2010,pammolli2002,riccaboni2010,riccaboni2013,chessa2013,nature}. Designed to keep track of the inter-industrial relationships, the input-output system is an ideal test bed for network science. Particularly the MRIO system can be viewed as an interdependent complex network, i.e., the WION, where the nodes are the individual industries in different economies and the edges are the flows between industries. 

Figure \ref{table1Net} provides a topological view of Table \ref{table_1}. The blue nodes are the individual industries. The red nodes are the value added sources from the two economies, whereas the green nodes are the final demand destinations in the two economies. The edges are with arrows indicating the directions of the monetary goods flows\footnote{Strictly speaking, the flows from the red nodes to the blue nodes are not goods but primary inputs in nature.} and with varying widths indicating the magnitudes of the flows. The color of the edge is set the same as the source node's. Finally, because we are only concerned with the inter-industrial input-output relationships, when formulating the WION, we focus our attention on the network among the blue nodes.   

\begin{table}[!t]
  \centering
  \caption{{\bf Top 20 industries identified by the three methods for selected years.} The first method is the final-demand-weighted backward linkage measure, $\boldsymbol{\mathrm{w}}$. The second is the PageRank centrality, $PR$. The third is the community coreness measure $|\mathrm{d}Q|$.}
  \resizebox{18cm}{!}{
    \begin{tabular}{@{\extracolsep{4pt}}cccccccccc@{}}
    \toprule
          & \multicolumn{3}{c}{\textbf{1995}} & \multicolumn{3}{c}{\textbf{2003}} & \multicolumn{3}{c}{\textbf{2011}} \\
    \cline{2-4} \cline{5-7} \cline{8-10}
    \textbf{Rank} & $\boldsymbol{\mathrm{w}}$     & $PR$    & $|\mathrm{d}Q|$    & $\boldsymbol{\mathrm{w}}$     & $PR$    & $|\mathrm{d}Q|$    & $\boldsymbol{\mathrm{w}}$     & $PR$    & $|\mathrm{d}Q|$ \\
    \textbf{1} & USA-Pub & USA-Pub & USA-Pub & USA-Pub & USA-Hth & USA-Obs & CHN-Cst & GBR-Hth & CHN-Cst \\
    \textbf{2} & JPN-Cst & USA-Tpt & JPN-Cst & USA-Hth & DEU-Tpt & USA-Est & USA-Pub & DEU-Tpt & USA-Obs \\
    \textbf{3} & USA-Cst & DEU-Tpt & USA-Obs & USA-Cst & USA-Pub & USA-Fin & USA-Hth & USA-Pub & CHN-Met \\
    \textbf{4} & USA-Hth & USA-Hth & USA-Cst & USA-Est & USA-Tpt & USA-Pub & USA-Est & CHN-Elc & USA-Pub \\
    \textbf{5} & USA-Est & DEU-Cst & USA-Est & USA-Rtl & GBR-Hth & USA-Hth & CHN-Elc & USA-Hth & USA-Est \\
    \textbf{6} & USA-Rtl & RUS-Hth & USA-Hth & CHN-Cst & ESP-Cst & CHN-Cst & USA-Rtl & CHN-Cst & CHN-Agr \\
    \textbf{7} & USA-Fod & DEU-Fod & JPN-Htl & JPN-Cst & DEU-Hth & JPN-Cst & USA-Cst & CHN-Met & CHN-Fod \\
    \textbf{8} & JPN-Pub & GBR-Cst & JPN-Met & USA-Fin & GBR-Cst & USA-Ocm & USA-Fin & USA-Tpt & USA-Fin \\
    \textbf{9} & USA-Tpt & USA-Cst & USA-Met & USA-Tpt & USA-Cst & CHN-Met & CHN-Fod & ESP-Cst & CHN-Min \\
    \textbf{10} & JPN-Est & FRA-Tpt & JPN-Obs & USA-Fod & USA-Obs & USA-Met & CHN-Mch & AUS-Cst & USA-Cok \\
    \textbf{11} & JPN-Hth & USA-Fod & DEU-Cst & USA-Htl & FRA-Tpt & JPN-Obs & JPN-Cst & ITA-Hth & CHN-Elc \\
    \textbf{12} & USA-Fin & GBR-Hth & JPN-Pub & USA-Ocm & TUR-Tex & JPN-Htl & USA-Fod & DEU-Hth & USA-Hth \\
    \textbf{13} & USA-Htl & USA-Obs & JPN-Hth & JPN-Pub & USA-Est & CHN-Agr & USA-Htl & USA-Obs & CHN-Omn \\
    \textbf{14} & JPN-Fod & JPN-Cst & JPN-Ocm & USA-Obs & AUS-Cst & USA-Cst & CHN-Tpt & RUS-Hth & CHN-Cok \\
    \textbf{15} & JPN-Rtl & DEU-Mch & JPN-Fod & JPN-Est & USA-Fod & JPN-Pub & JPN-Pub & CHN-Tpt & CHN-Mch \\
    \textbf{16} & DEU-Cst & ESP-Cst & JPN-Fin & JPN-Hth & ITA-Hth & USA-Agr & USA-Tpt & DEU-Mch & USA-Cst \\
    \textbf{17} & JPN-Elc & JPN-Tpt & USA-Agr & USA-Whl & DEU-Cst & USA-Tpt & USA-Ocm & FRA-Cst & CHN-Chm \\
    \textbf{18} & JPN-Whl & DEU-Met & USA-Fod & JPN-Tpt & DEU-Fod & JPN-Met & CHN-Pub & CHN-Tex & JPN-Obs \\
    \textbf{19} & JPN-Tpt & USA-Elc & JPN-Whl & CHN-Elc & DEU-Mch & JPN-Hth & USA-Obs & GBR-Cst & USA-Ocm \\
    \textbf{20} & JPN-Mch & USA-Est & USA-Pup & DEU-Tpt & CHN-Elc & CHN-Elc & USA-Whl & DEU-Met & JPN-Cst \\
    \botrule
    \end{tabular}%
    }
  \label{table_2}%
\end{table}%

\begin{figure}[ht]
  \begin{center}
    \subfigure[1995]{\label{WION1995}\includegraphics[width=8cm]{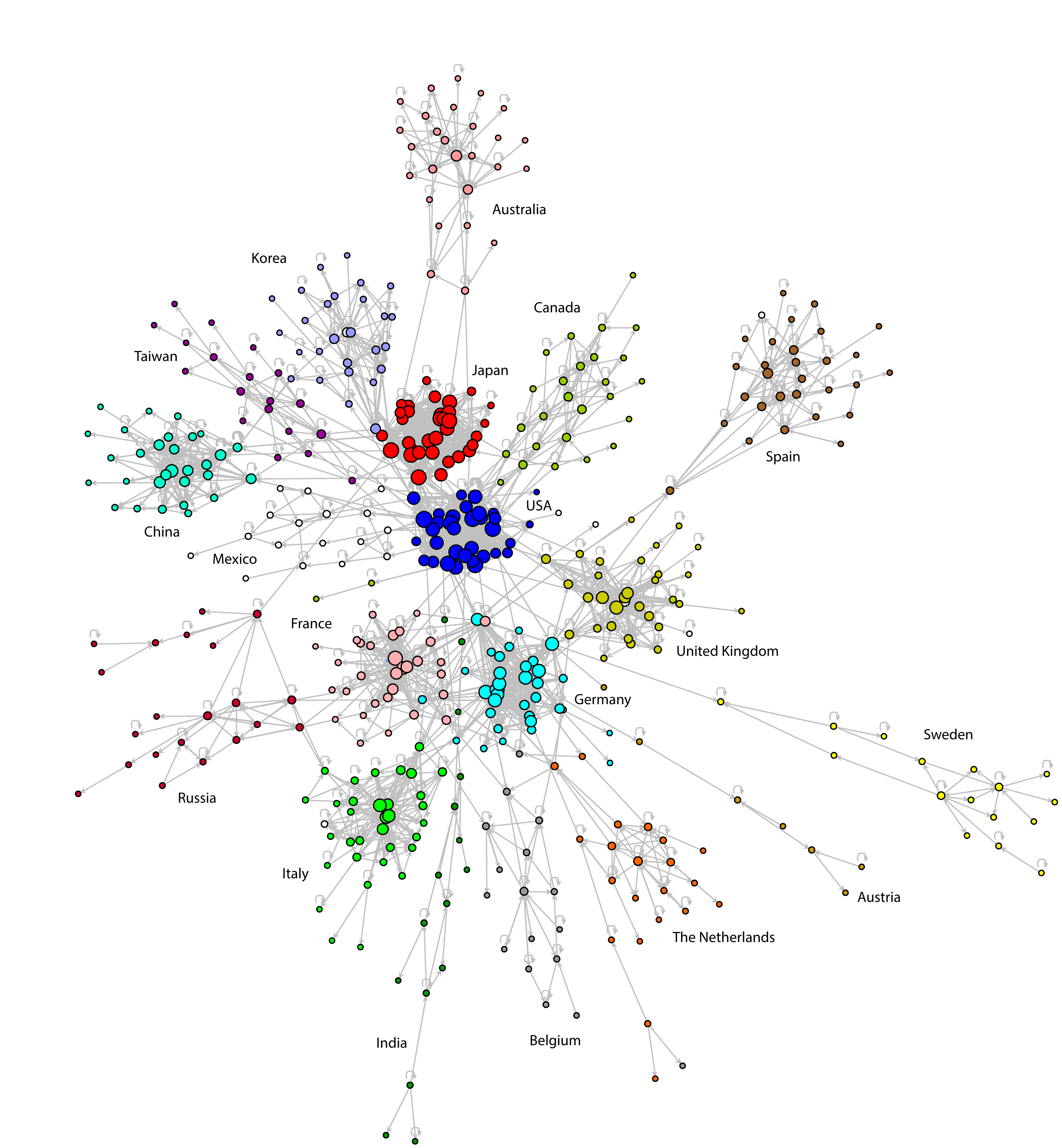}}
    \subfigure[2011]{\label{WION2011}\includegraphics[width=8cm]{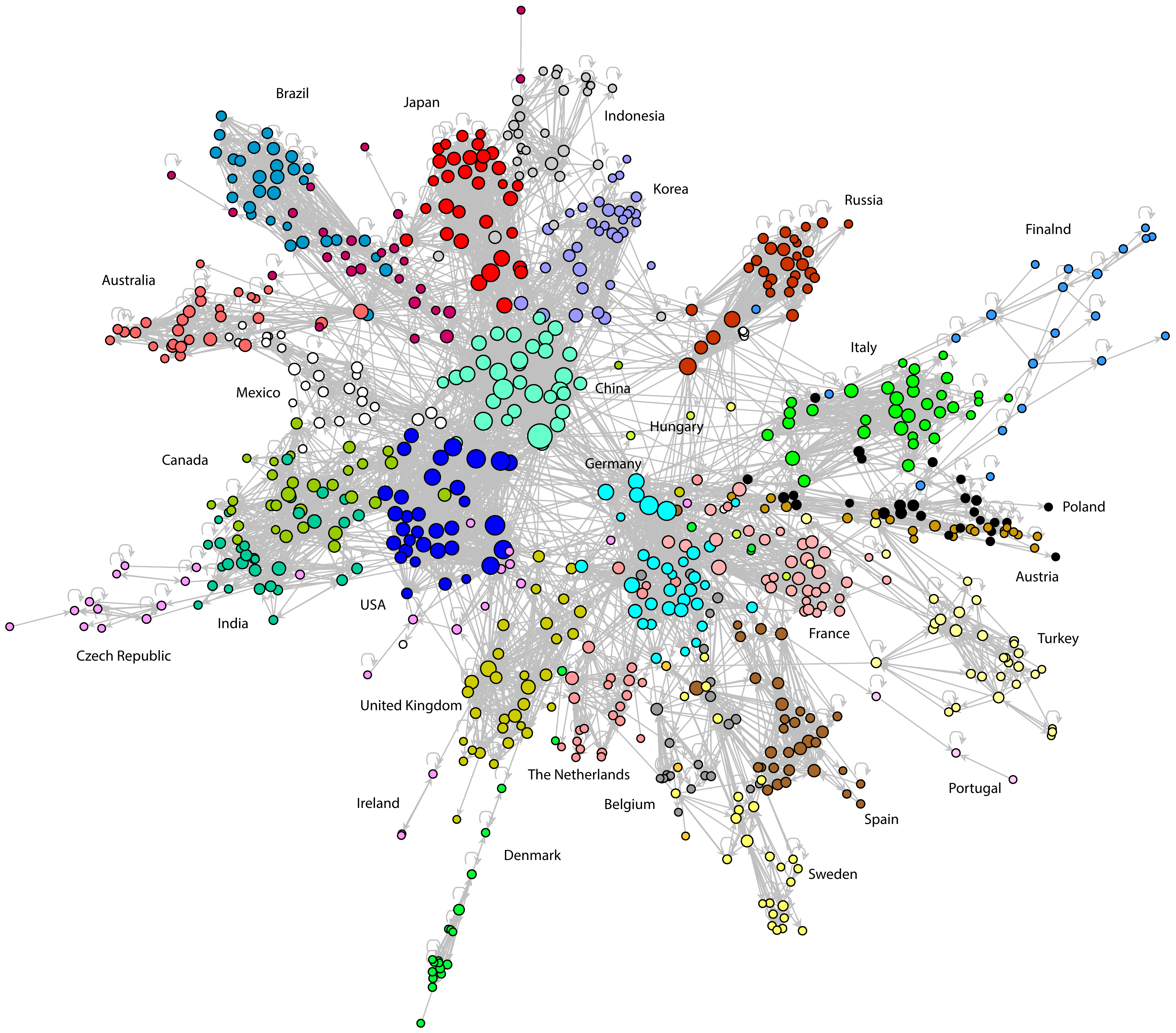}} \\
  \end{center}
  \caption{{\bf The WION in 1995 and 2011.} Each node represents a certain industry in a certain economy. The size of the node is proportional to its total degree (number of edges). The edges are directed and only those with strength greater than 1000 millions of US dollars are present. Finally, different colors represent different economies.}
  \label{WION9511}
\end{figure}

The visualization of the WION in 1995 (Figure \ref{WION1995}) and 2011 (Figure \ref{WION2011}) are shown in Figure \ref{WION9511}. Each node represents a certain industry in a certain economy. The size of the node is proportional to its total degree. The edges are directed and only those with strength greater than one billion US dollars are present. Finally, different colors represent different economies. Clearly the WION has become denser over time and some countries like China have moved to the core of the network, thus confirming the results in Table \ref{table_2}.  

In this section, we first examine some global network properties of the WION such as assortativity, clustering coefficient, and degree and strength distributions. We also study the subgraph structure and dynamics of the WION by using community detection techniques. Finally, we use the network-based PageRank centrality and community coreness measure to identify the key industries and economies in the WION and the results are different from the one obtained by the above Leontief-inverse-based method.  

\subsection{The Global Network Properties of the WION}

\begin{figure}[!t]
\centering
{\includegraphics[width=8cm]{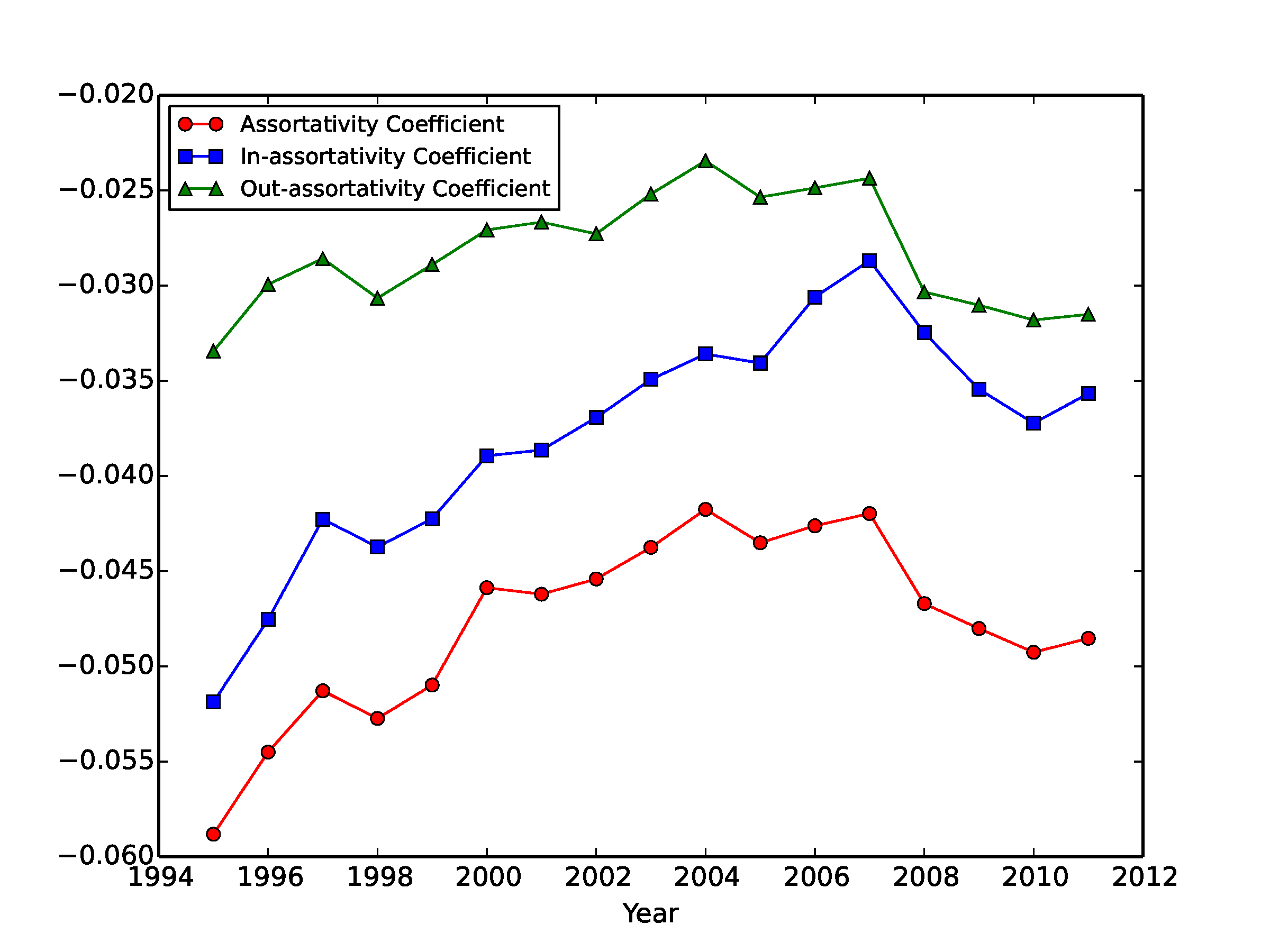}}
\caption{{\bf Assortativity of the WION over time.} From top to bottom, we show the over time out-degree assortativity, in-degree assortativity, and total-degree assortativity, respectively.} \label{assort}
\end{figure}

\begin{figure}[ht]
  \begin{center}
    \subfigure[All 40 Economies]{\label{foreign_all}\includegraphics[width=8cm]{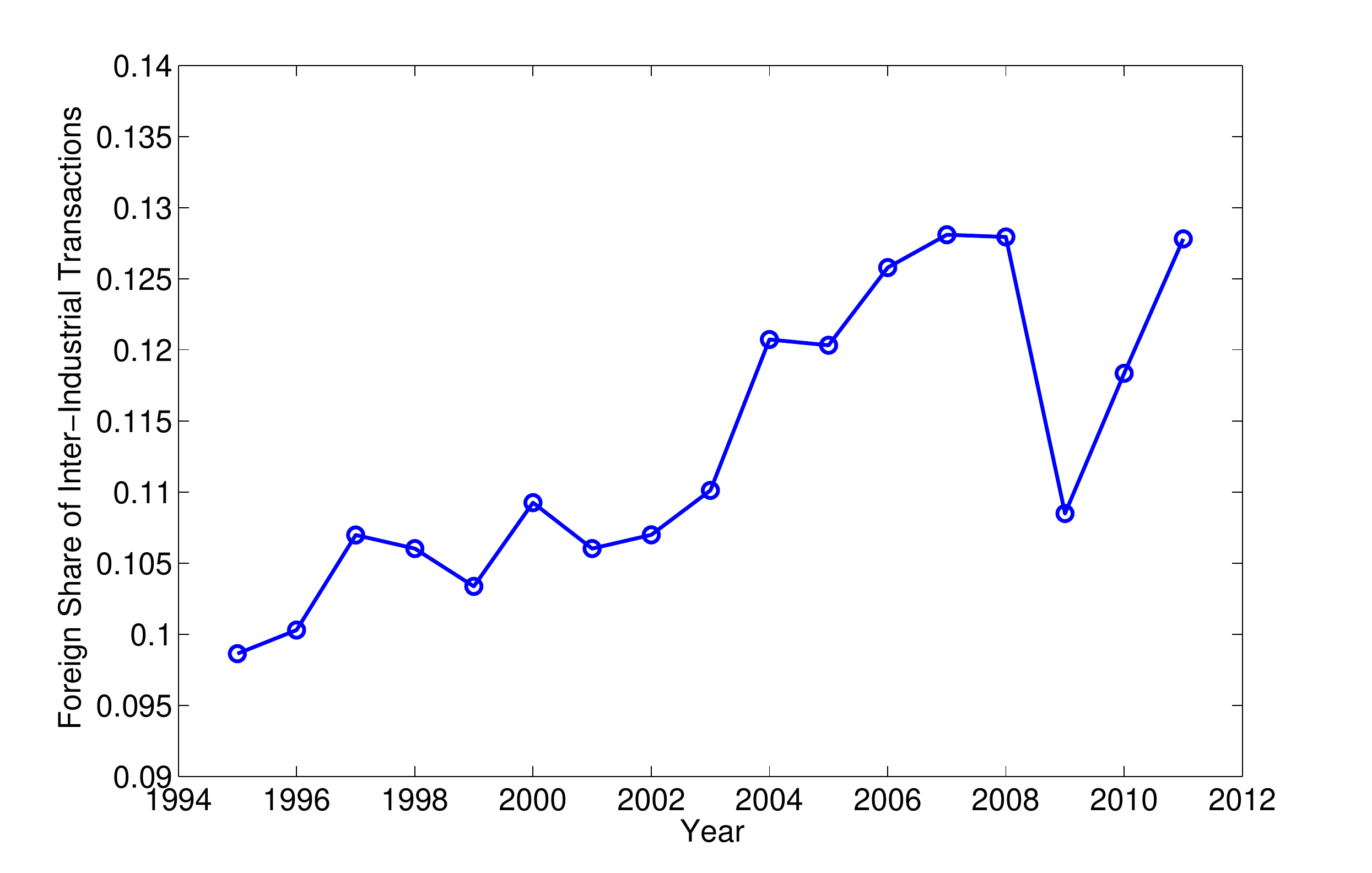}}
    \subfigure[Selected Regions]{\label{foreign_region}\includegraphics[width=8cm]{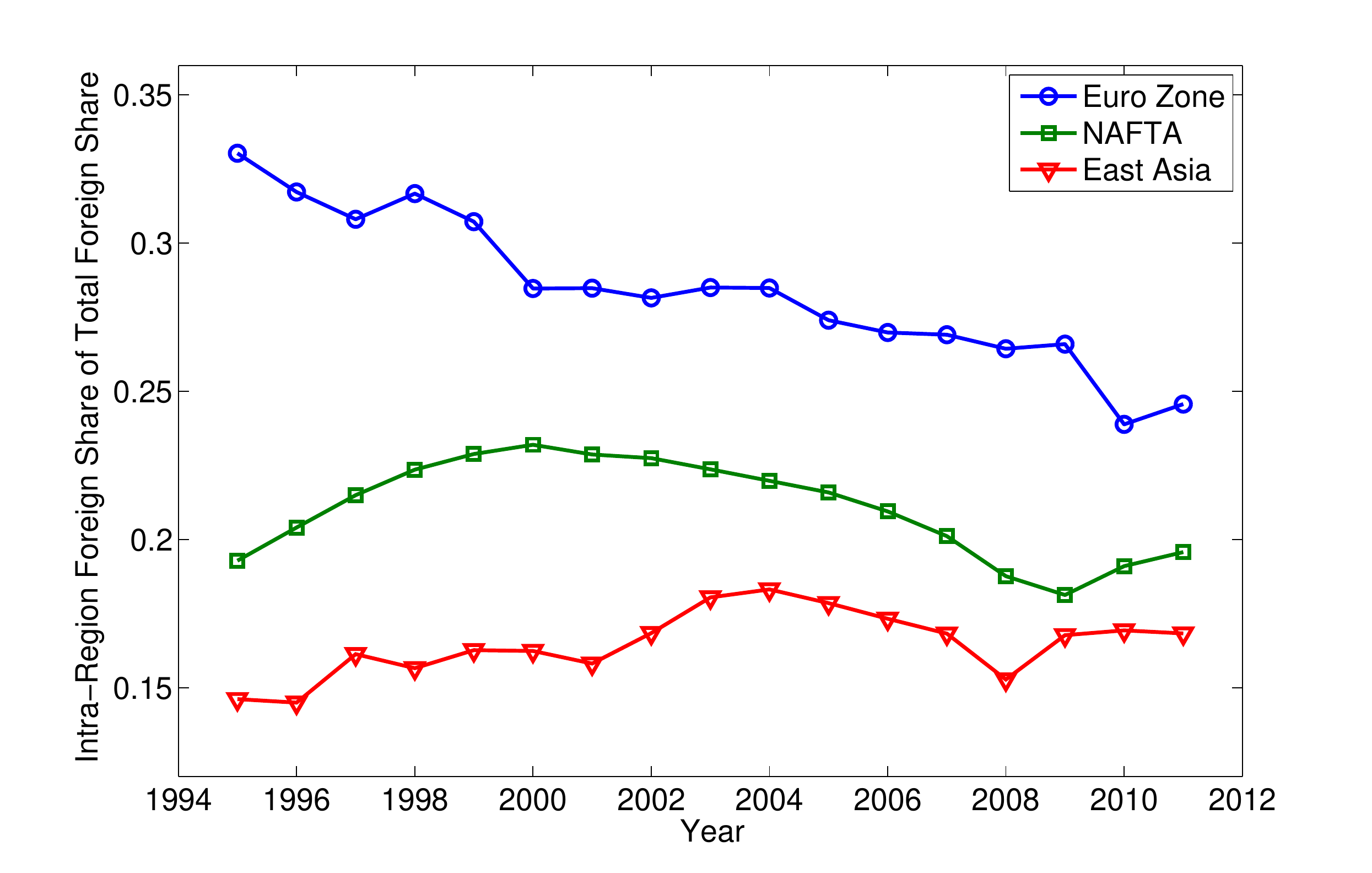}} \\
  \end{center}
  \caption{{\bf Globalized WION.} Figure \ref{foreign_all} shows the foreign share of the transactions matrix $\boldsymbol{\mathrm{Z}}$ over time. We calculate the percentage of inputs from foreign origins (or equivalently, the percentage of outputs to foreign destinations) of the transactions matrix $\boldsymbol{\mathrm{Z}}$ of the 40 WIOD economies. It can be viewed as a globalization indicator becasue it measures how much inter-industrial transactions are made through international trade. Figure \ref{foreign_region} considers the intra-region foreign share out of the total foreign share for some regions classified in Table \ref{table_A1} in the appendix. For the three regions, Euro Zone relies on the intra-region foreign trade the most and East Asia the least. Moreover, while the intra-region share in the other two regions fluctuates over time, it almost always declines in Euro Zone. Finally, all the three regions became less dependent on the intra-region foreign trade before the 2008 crisis. After the crisis, East Asia increased the intra-region foreign trade immediately, which is followed by NAFTA, and then by Euro Zone.}
  \label{foreign}
\end{figure}

Because the WION is directed, we can calculate the assortativity coefficient in three ways, namely, in-degree assortativity, out-degree assortativity, and total-degree assortativity. As shown in Figure \ref{assort}, they all behave similarly over time. First, they have all been negative throughout the whole period. Since assortativity measures the tendencies of nodes to connect with other nodes that have similar (or dissimilar) degrees as themselves, a negative coefficient means that dissimilar nodes are (slightly\footnote{In the case of WION, all the coefficients are of very small magnitude less than 0.06.}) more likely to be connected.\footnote{Notice that when calculating the assortativity for in-degree, out-degree, and total-degree, respectively, we consider the nodes as the neighbors of a given node if they are connected with the given node by only incoming edges, by only outgoing edges, and by either incoming or outgoing edges, respectively. In contrast, Carvalho \cite{carvalho2013survey} defines the neighborhood solely on the basis of the incoming edges and finds a positive assortative relationship.} One possible explanation of the negativity is that high-degree industries such as construction often take inputs (or supply outputs) from (or to) low-degree industries such as transport services. Moreover, the spatial constraints (each node has only few neighboring nodes in the same country) introduce degree-degree anticorrelations (disassortativity) since high degree sectors are in different countries and the probability to connect decays with distance \cite{havlin2014}. 
Second, all the coefficients show an increasing trend before 2007 and a significant decline after 2007. The behavior of the assortativity measures seems to be correlated with the trend of the foreign share in the inter-industrial transactions over time (Figure \ref{foreign_all}). That is, we can calculate a globalization indicator as the percentage of inputs from foreign origins (or equivalently, the percentage of outputs to foreign destinations) of the transactions matrix $\boldsymbol{\mathrm{Z}}$ of the 40 WIOD economies. Same as observed in assortativity, the foreign share of $\boldsymbol{\mathrm{Z}}$ had a steady growth (from 9.9\% in 1995 to 12.8\% in 2007) before 2007 and a sharp decrease after 2007\footnote{While the most severely depressed domestic edges during 2008-2009 in terms of the magnitude of the reduced flows are mostly within USA, the top 3 most impacted foreign edges are all from the mining industry to the coke and fuel industry and are from Canada to USA, from Netherlands to Belgium, and from Mexico to USA, respectively.}. The increase in the foreign share implies more interactions across economies and hence tends to make the WION less dissortative. The opposite happens when the foreign share goes down as a result of the global financial crisis.     
Third, we notice that the in-degree assortativity tends to be lower than the out-degree assortativity, but there is a tendency to close the gap between the two measures. We interpret this evidence as a clear signal of the globalization of production chains, that is to say, both global buying and selling hubs have now a higher chance to be connected across borders.

\begin{figure}[ht]
  \begin{center}
    \subfigure[Clustering Coefficient]{\label{cluster_all}\includegraphics[width=8cm]{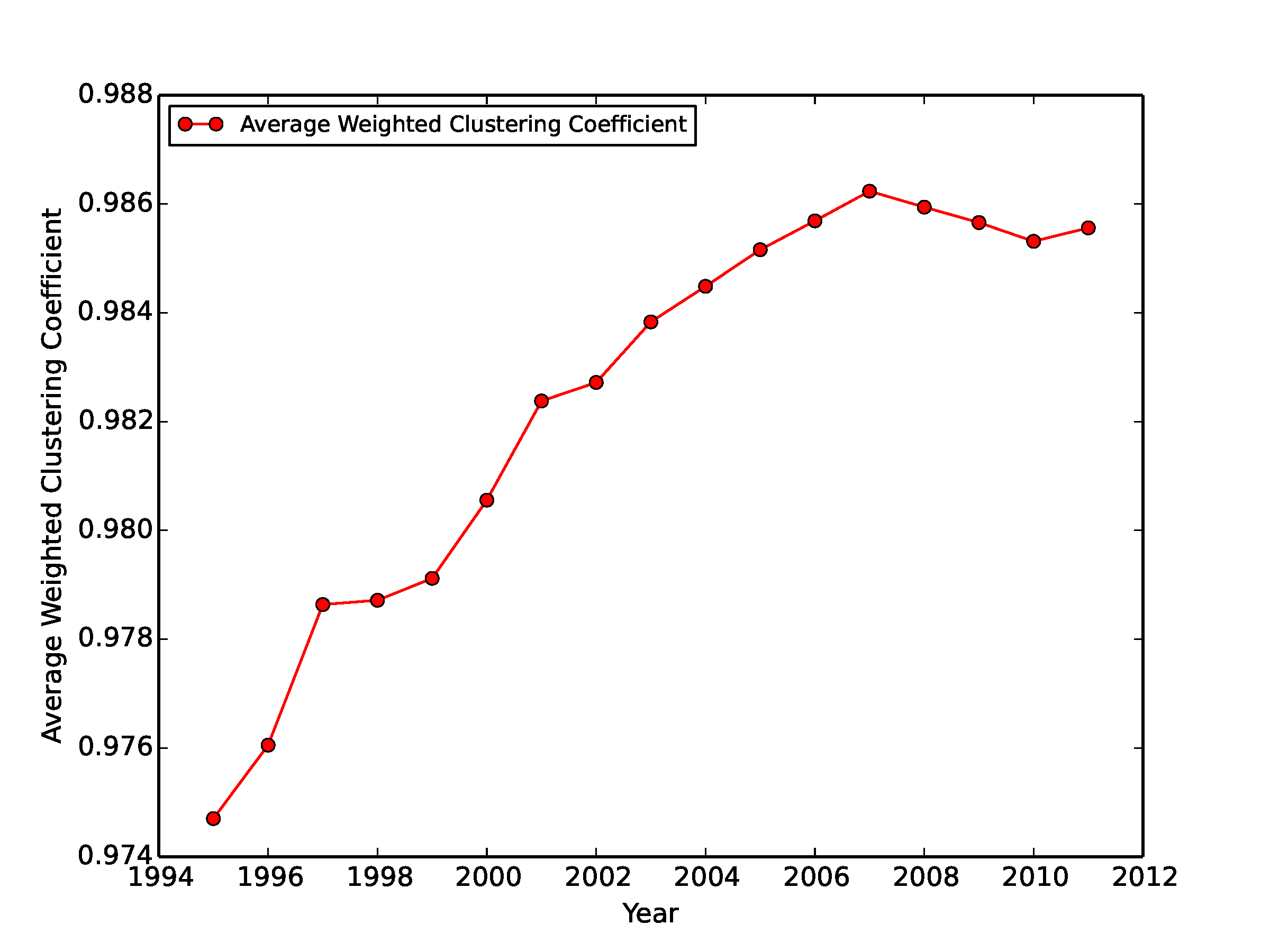}}
    \subfigure[Domestic and Foreign]{\label{cluster_fd}\includegraphics[width=8cm]{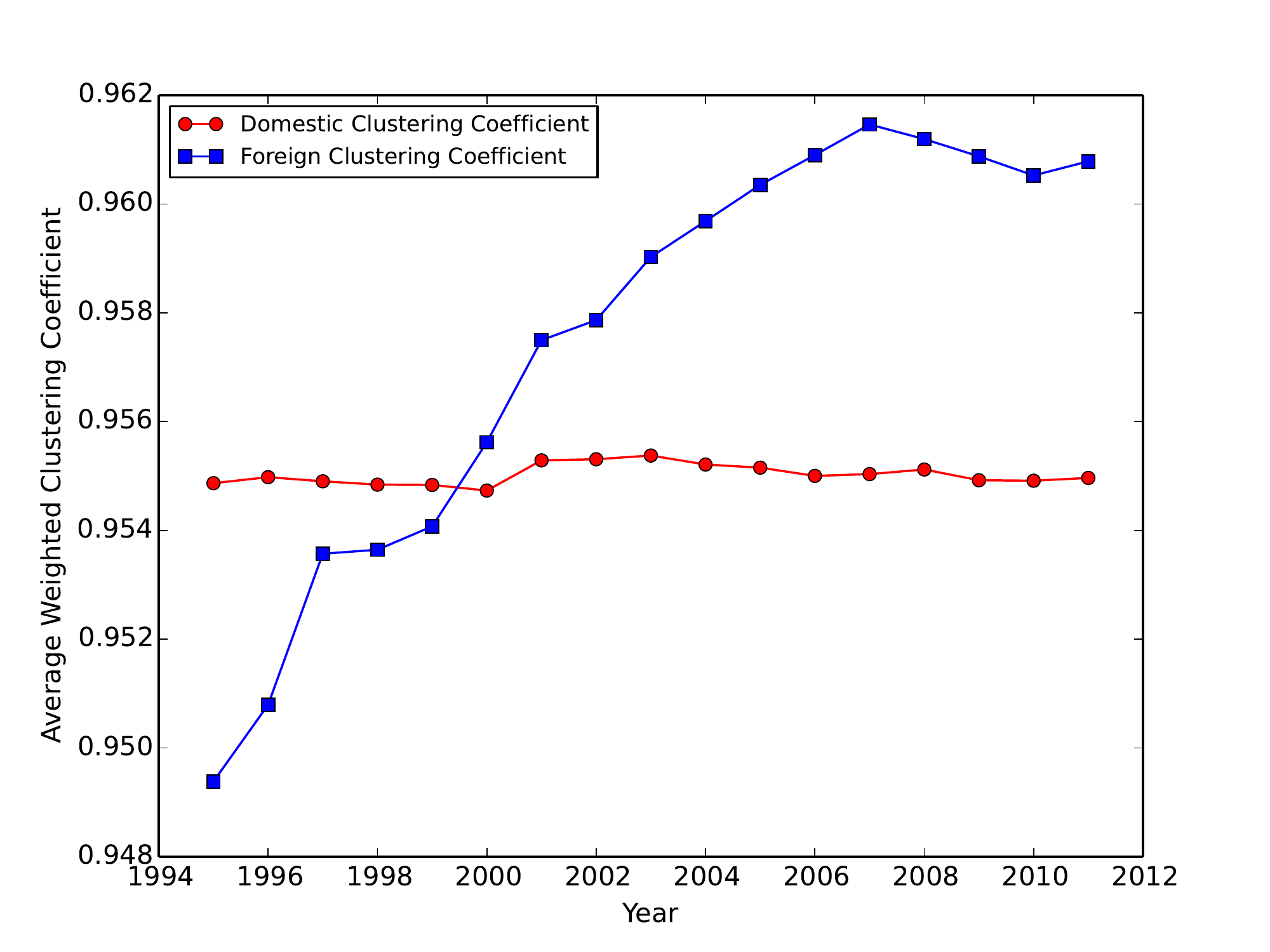}} \\
  \end{center}
  \caption{{\bf Clustering coefficient of the WION over time.} Figure \ref{cluster_all} shows the average weighted clustering coefficient of the WION over time. Figure \ref{cluster_fd} further decomposes the clustering coefficient into domestic clustering coefficient and foreign clustering coefficient. Clearly the behavior in Figure \ref{cluster_all} is more explained by the foreign clustering coefficient.}
  \label{cluster}
\end{figure}

The hump-shaped behavior is also observed in the clustering coefficient. Figure \ref{cluster_all} shows that the average weighted clustering coefficient of the WION has been steadily increasing but was followed by a decline since 2007. Again, a possible explanation is that the booming economy before 2007 introduced more interactions between industries, hence higher clustering coefficient, and the financial crisis after 2007 stifled the excess relationships.

We can also examine the global network properties of the WION by plotting its degree and strength distributions. Recall that the transactions matrix $\boldsymbol{\mathrm{Z}}$ is essentially a weighted adjacency matrix, which records the edge weights between any pair of nodes in the WION. If we denote the regular binary adjacency matrix as $\boldsymbol{\mathrm{D}}$, where $D_{ij}=D_{ji}=1$ if either $Z_{ij}>0$ or $Z_{ji}>0$, then we have the following definitions for a given node $i$: 1) In-degree: $D_i^{in}=\sum_{j\neq i}D_{ji}$; 2) Out-degree: $D_i^{out}=\sum_{j\neq i}D_{ij}$; 3) Total-degree: $D_i^{total}=D_i^{in}+D_i^{out}$; 4) In-strength: $S_i^{in}=\sum_{j\neq i}Z_{ji}$; 5) Out-strength: $S_i^{out}=\sum_{j\neq i}Z_{ij}$; 6) Total-strength: $S_i^{total}=S_i^{in}+S_i^{out}$. 

\begin{figure}[!t]
\centering
{\includegraphics[width=14cm]{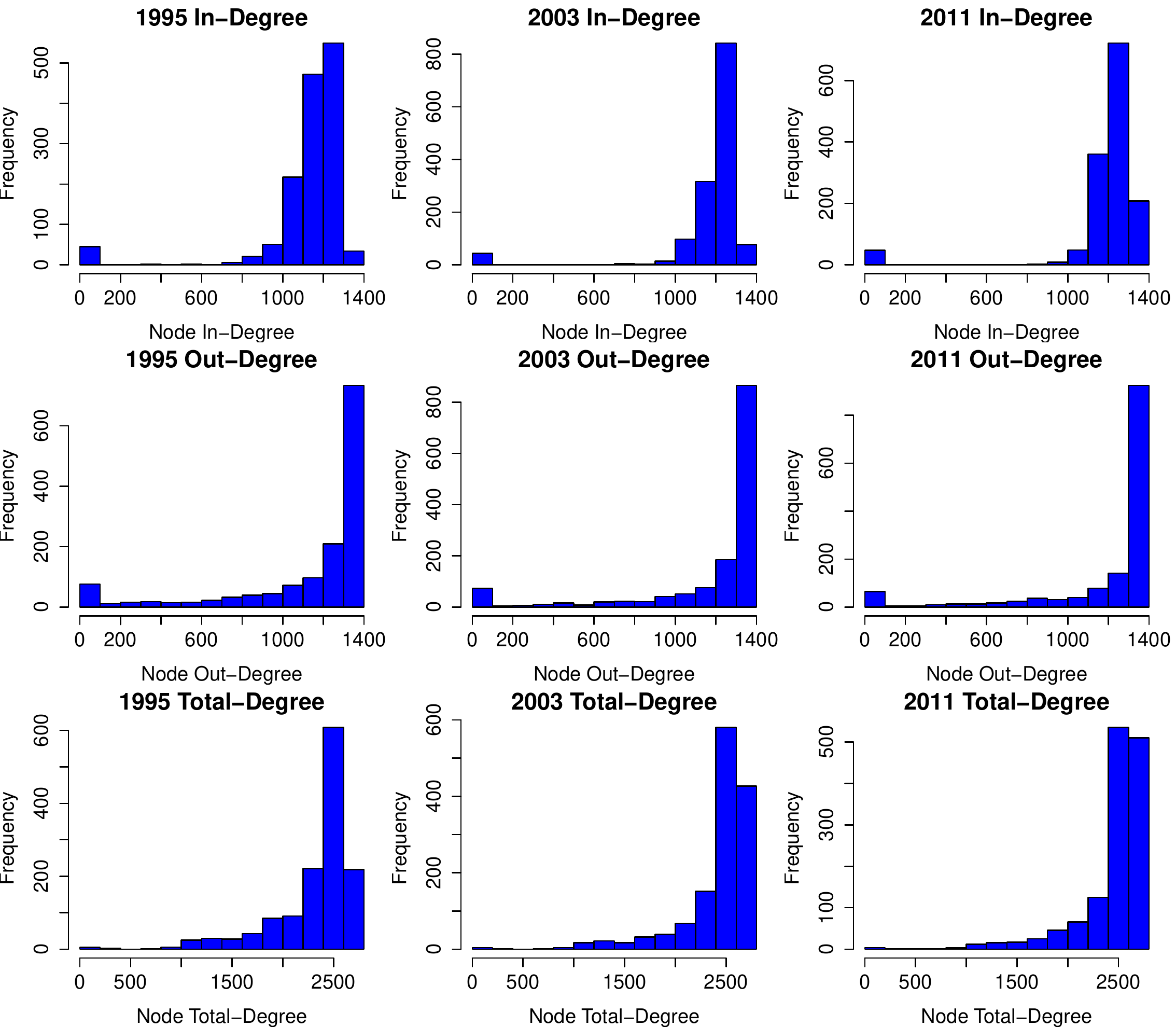}}
\caption{{\bf Histogram of in-degree, out-degree, and total-degree distributions for selected years.} For the selected years 1995, 2003, and 2011, the first row has the in-degree distributions while the second row and the third row have the out-degree and total-degree distributions respectively. The WION is characterized by the highly left-skewed degree distributions. Most nodes enjoy high-degree connections in the WION due to the aggregated industry classification.} \label{degreeDist}
\end{figure}

As shown in Figure \ref{degreeDist}, unlike other network systems such as the internet, where the degree distributions follow the power law, the WION is characterized by the highly left-skewed degree distributions. Most nodes enjoy high-degree connections in the WION because the industries are highly aggregated. That is, it is hard to find two completely disconnected industries given the high level of aggregation. Furthermore, the WION is almost complete, i.e., every node is connected with almost every node, if represented by unweighted edges\footnote{The same feature is also found in the input-output networks at the national level \cite{mcnerney2013network}. Using a single-year (2006) data of the WIOD, Carvalho \cite{carvalho2013survey} also reports the heavy-tailed but non-power-law degree distributions.}. 

\begin{figure}[!t]
\centering
{\includegraphics[width=14cm]{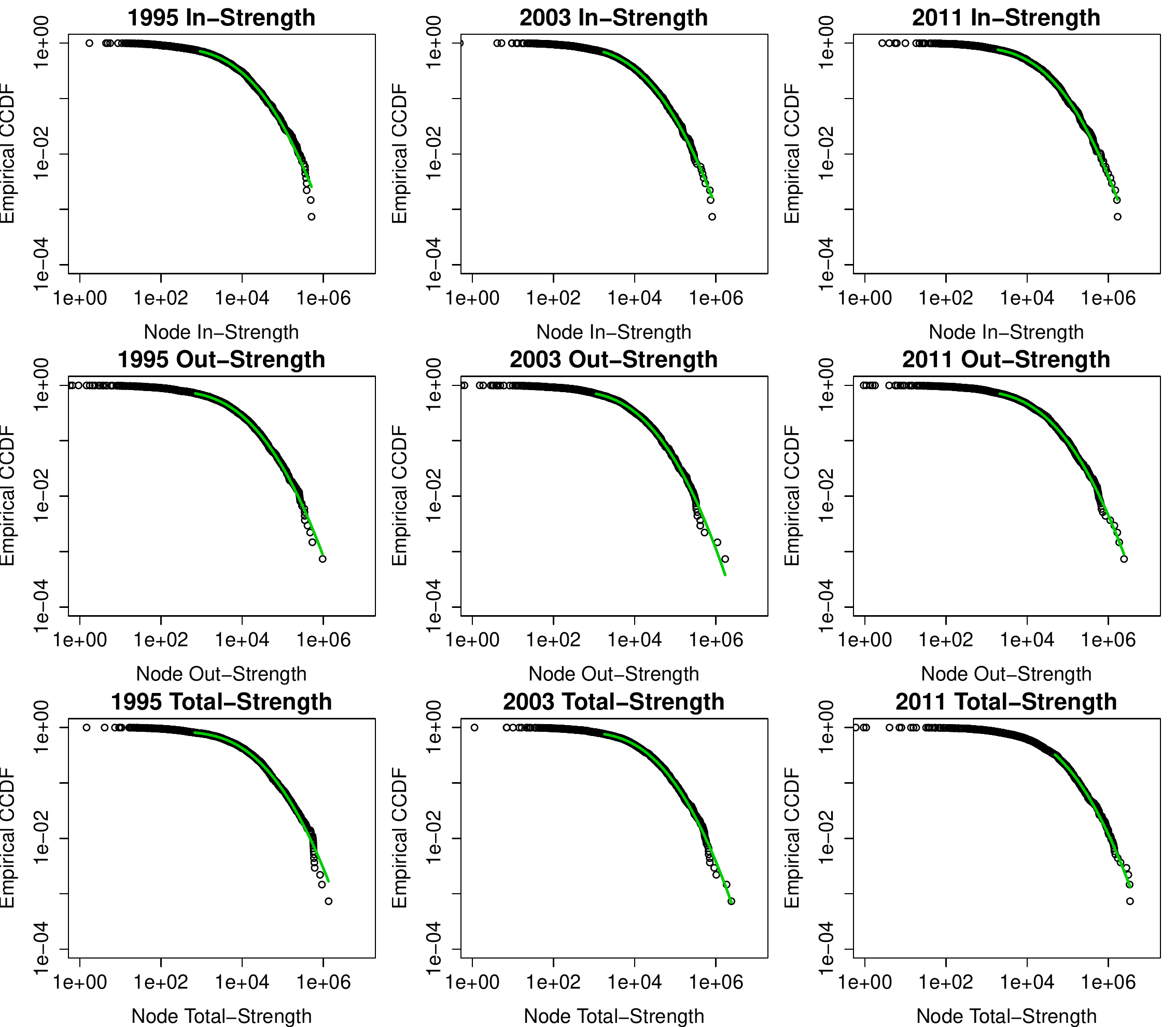}}
\caption{{\bf Empirical counter-cumulative distribution functions of in-strength, out-strength, and total-strength for selected years.} For the selected years 1995, 2003, and 2011, the first row has the in-strength distributions while the second row and the third row have the out-strength and total-strength distributions respectively. The observed data are in black circles while the green curve is the fitted log-normal distribution.} \label{strengthDist}
\end{figure}

We can also take into account the edge weights and examine the strength distributions of the WION. Figure \ref{strengthDist} shows the in-strength, out-strength, and total-strength distributions for the years 1995, 2003, and 2011. We perform Gabaix-Ibragimov test \cite{gabaix2009power,gabaix2011rank} to see if the tails of the distributions are Pareto but find no significant power-law tails. Moreover, like the previous studies at the national level \cite{mcnerney2013network}, the strength distributions can be well approximated by the log-normal distributions. As reasoned by Acemoglu et al. \cite{acemoglu2012network}, this asymmetric and heavy-tailed distribution of strength in the WION may serve as the origin of economic fluctuations. 

\subsection{The Community Detection in the WION}

Another main property of networks is the community structure, i.e. the partition of a network into clusters, with many edges connecting nodes in the same cluster 
and few connecting nodes between different ones \cite{fortunato2010}. In the following we use the modularity optimization method introduced by Newman and Girvan \cite{newman2004finding}. It is based on the idea that a random graph is not 
expected to have a community structure. Therefore, the possible existence of clusters is revealed by the comparison between the actual
density of edges in a subgraph and the expected density if the nodes are attached randomly. The
expected edge density depends on the chosen
null model, i.e., a copy of the original graph keeping some of its
structural properties but without community structure \cite{fortunato2010}.

The most popular null model, introduced by Newman and Girvan \cite{newman2004finding}, keeps the degree sequence and consists of a randomized version of the 
original graph, where edges are rewired at random, under the constraint that the expected degree of each node matches the degree of the node in the original graph. 

The modularity function to be optimized is, then, defined as:
\begin{equation}
Q = \frac{1}{2m} \sum_{ij} (A_{ij}-P_{ij})\delta({C_i,C_j}) 
\end{equation} \label{eq3}
where the summation operator runs over all the node pairs. $A$ is the adjacency matrix, and $m$ is the total number of edges. The $\delta$ function equals 1 if the two nodes $i$ and $j$ are in the same community and 0 otherwise. Finally, $P_{ij} = \frac{k_i k_j}{2m}$ is the probability of the presence of an edge between the two nodes $i$ and $j$ in the randomized null model.

\begin{figure}[!t]
\centering
{\includegraphics[width=14cm]{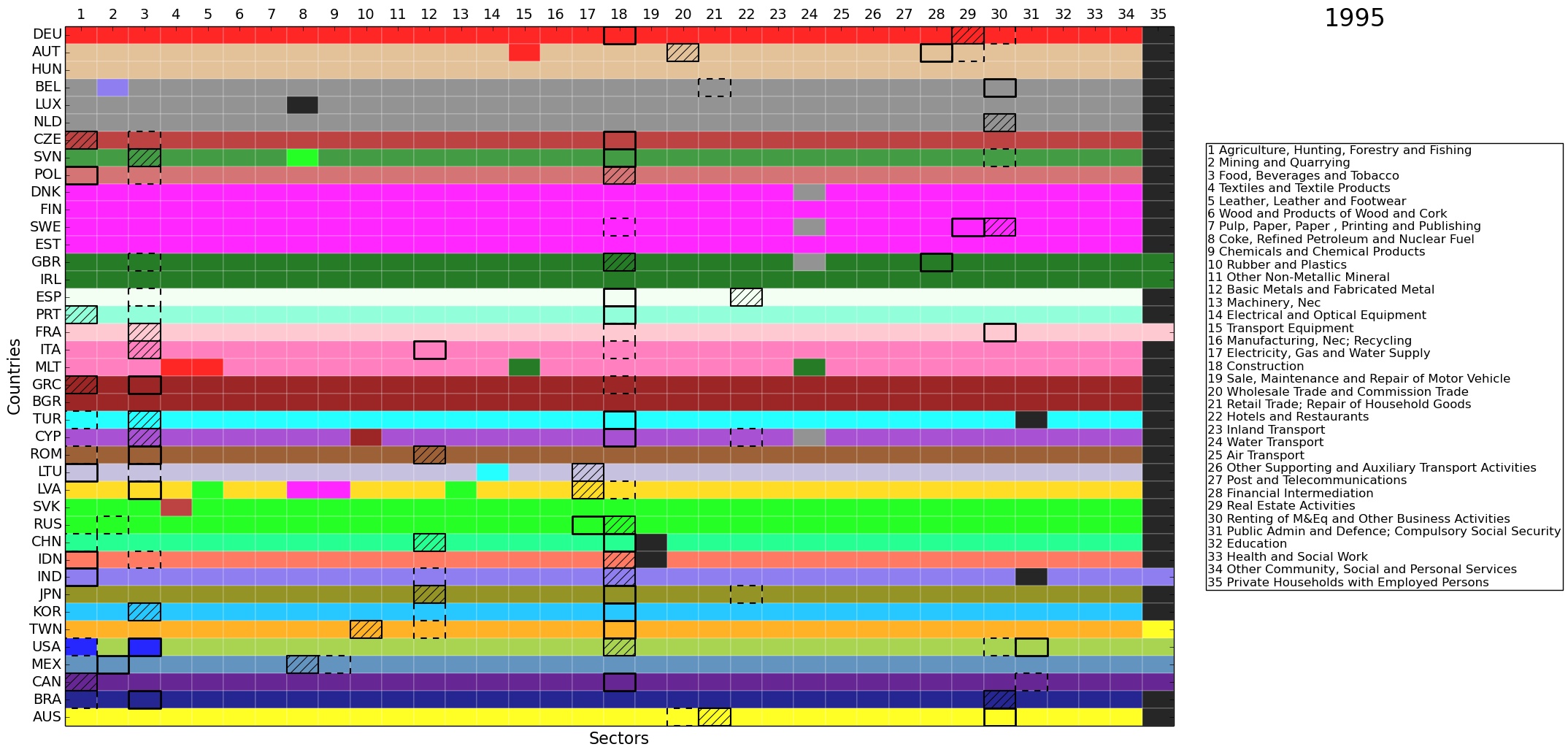}}
\caption{{\bf Community detection and community core detection results in 1995.} The economies are arranged by rows and the industries are arranged by columns. Each color represents a community detected, except that the black color indicates the isolated nodes with only self-loop. Within each community, the top 3 core economy-industry pairs are identified. The first place is with thick and solid border. The second place is with thick and dashed border. The third place is with border and texture.} \label{fig3}
\end{figure}

\begin{figure}[!t]
\centering
{\includegraphics[width=14cm]{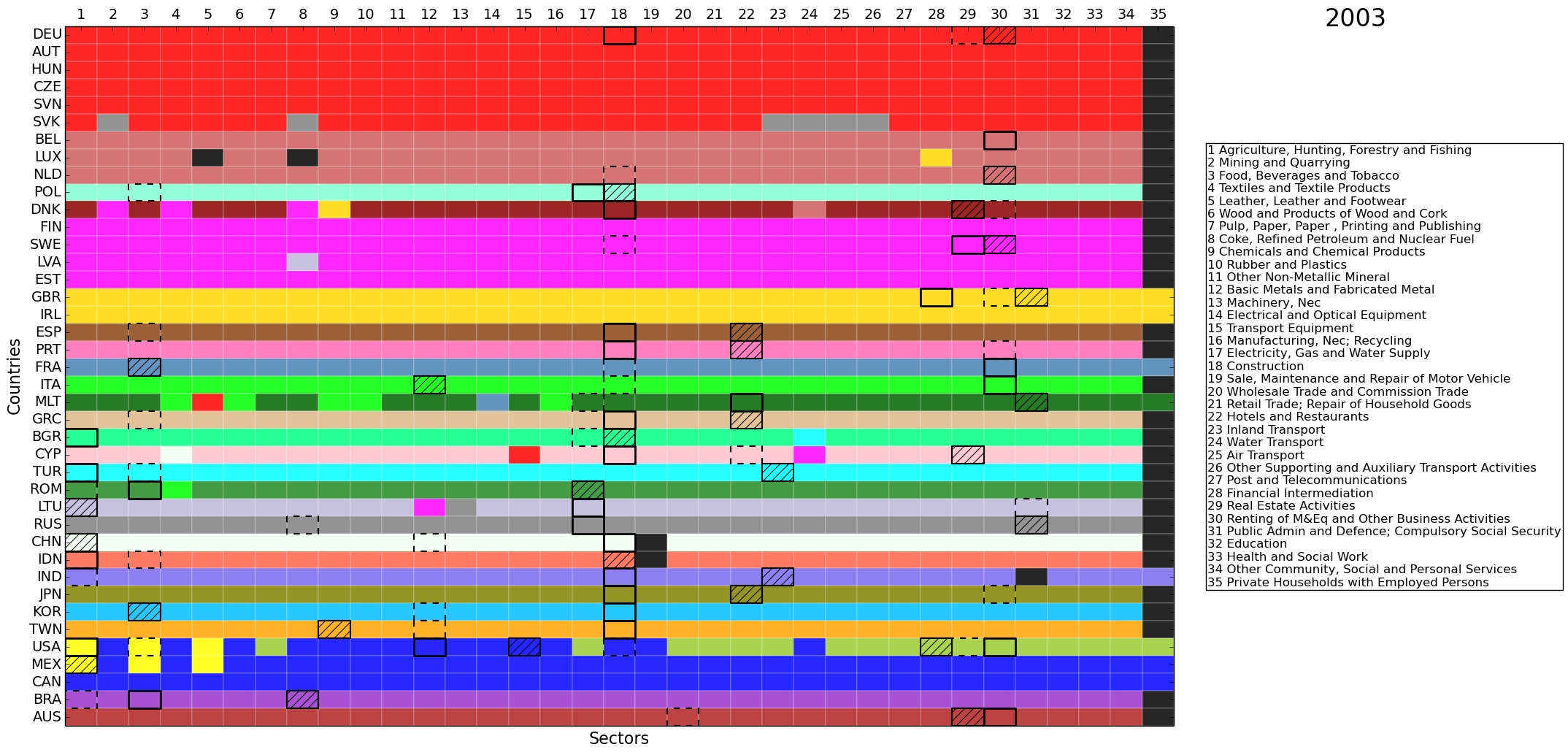}}
\caption{{\bf Community detection and community core detection results in 2003.} The economies are arranged by rows and the industries are arranged by columns. Each color represents a community detected, except that the black color indicates the isolated nodes with only self-loop. Within each community, the top 3 core economy-industry pairs are identified. The first place is with thick and solid border. The second place is with thick and dashed border. The third place is with border and texture.} \label{fig4}
\end{figure}

\begin{figure}[!t]
\centering
{\includegraphics[width=14cm]{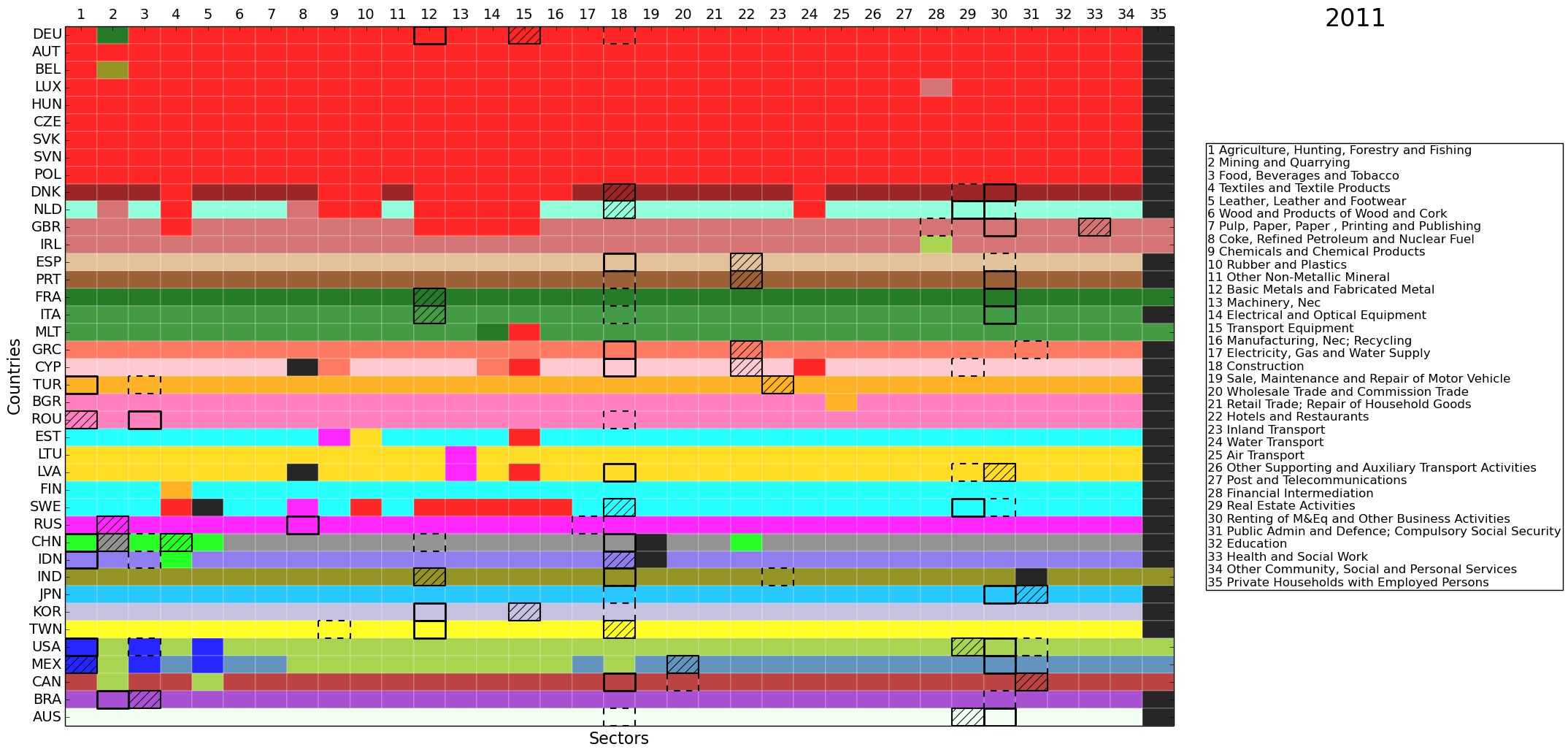}}
\caption{{\bf Community detection and community core detection results in 2011.} The economies are arranged by rows and the industries are arranged by columns. Each color represents a community detected, except that the black color indicates the isolated nodes with only self-loop. Within each community, the top 3 core economy-industry pairs are identified. The first place is with thick and solid border. The second place is with thick and dashed border. The third place is with border and texture.} \label{fig5}
\end{figure}

Figures \ref{fig3}, \ref{fig4}, and \ref{fig5} report the community detection results for the selected years 1995, 2003, and 2011, respectively\footnote{We perform the community detection for all available years (1995-2011). Results are available upon request.}. The 40 countries in the WIOD are arranged by rows while the 35 industries are arranged by columns. Different colors indicate different communities detected. There are two interesting findings in our results. First, most communities were based on a single economy, i.e., the same color often goes through a single row. This echoes one of the features of the WION mentioned in Section \ref{sec:intro}, i.e., most of the inter-industrial activities are still restricted in the country border. Second, for all the three years selected, we always color the community involving Germany in red and put it on the top. As a result, our algorithm captures a growing Germany-centered\footnote{It is centered on Germany because the community core detection results below show that the cores of this red community are all within Germany.} input-output community. 

Since the WIOD monetary goods flows are based on undeflated current prices, one possible reason for the emergence of the German community is that the community members may have experienced significantly more inflation and/or exchange rate volatility than other regions in the world. Referring to the World Bank inflation data and the exchange rate data used in the WIOD, we show that this is hardly the case. Panel (a) of Figure \ref{inflation_exchange} in the appendix compares the average inflation rate, i.e., the annual GDP deflator, across all the WIOD economies\footnote{The data is unavailable for Taiwan. We also exclude Bulgaria because it had a hyperinflation in 1997, which will bias the average if included. The data source is the World Development Indicators, the World Bank, \url{http://data.worldbank.org/indicator/NY.GDP.DEFL.KD.ZG}.} with the average annual GDP deflator across the 9 major member economies in the German community detected in 2011, i.e., Germany, Austria, Belgium, Luxembourg, Hungary, Czech Republic, Slovakia, Slovenia, and Poland. During 1995-2011, the average inflation of the German community was almost always below that of all the WIOD economies. Panel (b) of Figure \ref{inflation_exchange} in the appendix compares the average exchange rate, i.e., US dollars per unit of local currency, across all the WIOD economies\footnote{The data source is the exchange rate data used in the WIOD, \url{http://www.wiod.org/protected3/data/update_sep12/EXR_WIOD_Sep12.xlsx}.} with the average exchange rate across the above 9 major economies in the German community detected in 2011. The average exchange rate of the German community was basically below that of all the WIOD economies before 2000. Only from 2001, the community average became slightly (no more than 16\%) higher than the overall average. Therefore, the emergence of the German community cannot be attributed to inflation or exchange rate dynamics. Since most of the 40 economies in the WIOD are in Europe, we cannot rule out the possibility that similar regional input-output communities are emerging in other continents. Indeed, we find also an integrated NAFTA community in North America. However, since many Asian economies are not included in the WIOD, we cannot argue if a similar trend is ongoing in the Far East.

Within each community, we also carry out the community core detection (see below for more technical details). In Figures \ref{fig3}-\ref{fig5}, we identify the top 3 core economy-industry pairs for each community. The first place is with thick and solid border. The second place is with thick and dashed border. The third place is with border and texture. In general, the cores are mostly concentrated in the industries of agriculture (1), mining (2), food (3), metals (12), construction (18), and financial, business, and public services (28-31). Over time, while the services industries (28-31) have become the cores in more and more developed economies, the primary industries (1-3) have become less central in the developed economies and have only remained as the cores in a few emerging economies, which is consistent with the Kuznets facts \cite{kuznets1957quantitative,kuznets1973modern}. Furthermore, for the growing community centered on Germany, the cores are always identified in Germany (that is why we simply call it the German community) for the three selected years. It is also worth noting that, the German industry of transport equipment (15) is identified as a core in 2011 and the car industry is the most integrated in the German community, which spans over 17 economies.

\subsection{The Network-Based Methods of Identifying the Key Industries}

Since on a global scale the traditional assumption of stable input-output technical coefficients is violated due to the dynamics of international trade, the traditional final-demand-weighted backward linkage measure alone is insufficient to evaluate the importance of any given industry on the global economy. However, the networks approach provides us a holistic view of the global production system and we can compute various centrality measures to compare the nodes in the network.  Here we focus on two network-based methods of identifying the key industries in the WION, PageRank centrality\footnote{We choose PageRank over other centrality measures such as closeness and betweenness because the former systematically measures the influence of a given node and has been widely used in the previous literature to identify the key nodes \cite{acemoglu2012network,carvalho2013survey}.} and community coreness measure.  

\subsubsection{PageRank Centrality}

Given a network, it is a problem of capital importance to bring order to its structure by ranking nodes according to their relevance. Among the many proposed, a successful and widely used centrality measure is PageRank \cite{page1999pagerank}, a Google patented method. The idea is that the nodes are considered important if they are connected by other important nodes.  

Since the WION is weighted, we use a weighted version of PageRank, which is computed iteratively as follows:
\begin{enumerate}
\item At $t=0$, an initial probability distribution is assumed, usually  $PR(i; 0) = \frac{1}{N}$ where $N$ is the total number of nodes;
\item At each time step, the PageRank of node $i$ is computed as:
\begin{equation}
PR(i;t+1) = \frac{1-d}{N} + d \sum_{j \in M(i)} \frac{PR(j; t)w_{ij}}{S(j)}
\end{equation}
where $M(i)$ are the in-neighbors of $i$, $w_{ij}$ is the weight of the link between the nodes $i$ and $j$, $S$ is the sum of the weights of the outgoing edges from $j$, and the damping factor $d$ is set to its default value, 0.85.
\end{enumerate}

In Table \ref{table_2}, the second column of each year is produced by the PageRank centrality, which is denoted by $PR$.\footnote{Our PageRank result differs from the one reported by Carvalho \cite{carvalho2013survey}, where he uses an unweighted version of PageRank.} Unlike the final-demand-weighted backward linkage measure, where only 4 economies are among the top 20, the PageRank centrality recognizes 10 economies in the top 20 list for the three selected years. 

Tables \ref{table_A5} and \ref{table_A6} in the appendix provide an alternative way of viewing the key industries and economies over time identified by the PageRank centrality. In particular, Table \ref{table_A5} lists the most important economies by industry while Table \ref{table_A6} lists the most important industries by economy.

\subsubsection{Community Coreness Measure}

The other network-based method of identifying the key industries is the community coreness measure. Nodes of a community
do not have the same importance for the community stability: the removal of a node in the core of the community
affects the partition much more than the deletion of a node that stays on the periphery of the community \cite{deleo2013}. Therefore, in the following we 
define a novel way of detecting cores inside communities by using the properties of the modularity
function \ref{eq3}.

By definition, if the modularity associated with a network has been optimized, every perturbation in the
partition leads to a negative variation in the modularity, $\mathrm{d}Q$. If we move a node from its community,
we have $M-1$ possible choices, with $M$ as the number of communities, as the node's new host community. It is possible to define 
the $|\mathrm{d}Q|$ associated with each node as the smallest variation in absolute value (or the closest to 0 since $\mathrm{d}Q$ is always 
a negative number) of all the possible choices. We call $|\mathrm{d}Q|$ the community coreness measure. 

In the WION, once we have the $|\mathrm{d}Q|$ for each industry, we can consider the one with the biggest $|\mathrm{d}Q|$ the most important. We can also normalize the $|\mathrm{d}Q|$ to identify the most important nodes within each community. The results are shown in Figures \ref{fig3}, \ref{fig4}, and \ref{fig5}, where the first place in each community is with thick and solid border, the second place is with thick dashed border, and the third place is with both border and texture.

In Table \ref{table_2}, the third column of each year is produced by the community coreness measure, which is denoted again by $|\mathrm{d}Q|$. Interestingly, like the final-demand-weighted backward linkage measure, the community coreness measure also only includes China, Germany, Japan, and USA in the top 20 list for the selected years.  

Tables \ref{table_A7} and \ref{table_A8} in the appendix provide an alternative way of viewing the key industries and economies over time identified by the community coreness measure. In particular, Table \ref{table_A7} lists the most important economies by industry while Table \ref{table_A8} lists the most important industries by economy. 

Now we have totally three methods to identify the key industries in the WION, the traditional final-demand-weighted backward linkage measure, the PageRank centrality measure, and the community coreness measure. They have different results from each other. For instance, the industry of transport equipment in Germany is captured by the PageRank but not by the other two while the industry of other business activities in USA is more important by $|\mathrm{d}Q|$ than by the other two (see Table \ref{table_2}). Table \ref{table_3} reports the correlation coefficient matrix among the three methods for the selected years 1995, 2003, and 2011. We find that all the three methods are positively correlated, while $\boldsymbol{\mathrm{w}}$ and $|\mathrm{d}Q|$ are correlated even more. Therefore, the network-based $|\mathrm{d}Q|$ and especially $PR$ can be used to complement, if not to substitute, $\boldsymbol{\mathrm{w}}$ to identify the key industries in the WION. 

\begin{table}
  \centering
  \caption{{\bf Correlation coefficient matrix among the three key-industry-identification methods for selected years.} The first method is the final-demand-weighted backward linkage measure, $\boldsymbol{\mathrm{w}}$. The second is the PageRank centrality, $PR$. The third is the community coreness measure $|\mathrm{d}Q|$.}
  \resizebox{18cm}{!}{
    \begin{tabular}{@{\extracolsep{4pt}}cccccccccccc@{}}
    \toprule
    \multicolumn{4}{c}{\textbf{1995}} & \multicolumn{4}{c}{\textbf{2003}} & \multicolumn{4}{c}{\textbf{2011}} \\
    \cline{1-4} \cline{5-8} \cline{9-12}
          & $\boldsymbol{\mathrm{w}}$     & $PR$    & $|\mathrm{d}Q|$    &       & $\boldsymbol{\mathrm{w}}$     & $PR$    & $|\mathrm{d}Q|$    &       & $\boldsymbol{\mathrm{w}}$     & $PR$    & $|\mathrm{d}Q|$ \\
    $\boldsymbol{\mathrm{w}}$     & 1     & 0.664224 & 0.819625 & $\boldsymbol{\mathrm{w}}$     & 1     & 0.688819 & 0.724121 & $\boldsymbol{\mathrm{w}}$     & 1     & 0.64281 & 0.754442 \\
    $PR$    & 0.664224 & 1     & 0.650459 & $PR$    & 0.688819 & 1     & 0.596233 & $PR$    & 0.64281 & 1     & 0.592057 \\
    $|\mathrm{d}Q|$    & 0.819625 & 0.650459 & 1     & $|\mathrm{d}Q|$    & 0.724121 & 0.596233 & 1     & $|\mathrm{d}Q|$    & 0.754442 & 0.592057 & 1 \\
    \botrule
    \end{tabular}%
    }
  \label{table_3}%
\end{table}%

\section{Concluding Remarks} \label{sec:conclusion}

This paper investigates a MRIO system characterized by the recently available WIOD database. By viewing the world input-output system as an interdependent network where the nodes are the individual industries in different economies and the edges are the monetary goods flows between industries, we study the network properties of the so-called world input-output network (WION) and document its evolution over time. We are able to quantify not only some global network properties such as assortativity, clustering coefficient, and degree and strength distributions, but also its subgraph structure and dynamics by using community detection techniques. Over time, we trace the effects of globalization and the 2008-2009 financial crisis. We notice that national economies are increasingly interconnected in global production chains. Moreover, we detect the emergence of regional input-output community. In particular we see the formation of a large European community led by Germany. Finally, because on a global scale the traditional assumption of stable input-output technical coefficients is violated due to the dynamics of international trade, we also use the network-based PageRank centrality and community coreness measure to identify the key industries in the WION and the results are different from the one obtained by the traditional final-demand-weighted backward linkage measure.

As mentioned above, due to the limited coverage of the WIOD, we cannot argue if the input-output integration is also observed in other continents. Therefore, in our future work, we will utilize another MRIO database, EORA \cite{lenzen2012mapping,lenzen2013building}, which covers about 187 countries in the world and the years from 1990 to 2011. Moreover, since each of the three methods of identifying the key industries captures a different aspect of the importance of any given industry, future work is also needed to compare the methods so as to identify the systematically important industries for the global economy.

\section{Acknowledgments}

Authors thank Michelangelo Puliga for insightful discussions. All authors acknowledge support from the FET projects MULTIPLEX 317532 and SIMPOL 610704 and the PNR project CRISIS Lab. MR and ZZ acknowledge funding from the MIUR (FIRB project RBFR12BA3Y). FC gratefully acknowledges Sardinia Regional Government for the financial support of her PhD scholarship (P.O.R. Sardegna F.S.E. Operational Programme of the Autonomous Region of Sardinia, European Social Fund 2007–2013 - Axis IV Human Resources, Objective l.3, Line of Activity l.3.1.).

\bibliographystyle{unsrt}
\bibliography{WION}

\newpage \clearpage

\appendix*

\setcounter{table}{0}
\renewcommand\thetable{A\arabic{table}}

\setcounter{figure}{0}
\renewcommand\thefigure{A\arabic{figure}}

\begin{center}\textbf{APPENDIX}\end{center}

\begin{table}[H] 
  \centering
  \caption{{\bf List of WIOD economies.}}
  \resizebox{18cm}{!}{
    \begin{tabular}{@{\extracolsep{4pt}}llllllllll@{}}
    \toprule
    \multicolumn{2}{c}{\textbf{Euro-Zone}} & \multicolumn{2}{c}{\textbf{Non-Euro EU}} & \multicolumn{2}{c}{\textbf{NAFTA}} & \multicolumn{2}{c}{\textbf{East Asia}} & \multicolumn{2}{c}{\textbf{BRIIAT}} \\
    \cline{1-2} \cline{3-4} \cline{5-6}  \cline{7-8} \cline{9-10}
    \textbf{Economy} & \textbf{3L Code} & \textbf{Economy} & \textbf{3L Code} & \textbf{Economy} & \textbf{3L Code} & \textbf{Economy} & \textbf{3L Code} & \textbf{Economy} & \textbf{3L Code} \\
    
    Austria & AUT   & Bulgaria & BGR   & Canada & CAN   & China & CHN   & Australia & AUS \\
    Belgium & BEL   & Czech Rep. & CZE   & Mexico & MEX   & Japan & JPN   & Brazil & BRA \\
    Cyprus & CYP   & Denmark & DNK   & USA   & USA   & South Korea & KOR   & India & IND \\
    Estonia & EST   & Hungary & HUN   &       &       & Taiwan & TWN   & Indonesia & IDN \\
    Finland & FIN   & Latvia & LVA   &       &       &       &       & Russia & RUS \\
    France & FRA   & Lithuania & LTU   &       &       &       &       & Turkey & TUR \\
    Germany & DEU   & Poland & POL   &       &       &       &       &       &  \\
    Greece & GRC   & Romania & ROM   &       &       &       &       &       &  \\
    Ireland & IRL   & Sweden & SWE   &       &       &       &       &       &  \\
    Italy & ITA   & UK    & GBR   &       &       &       &       &       &  \\
    Luxembourg & LUX   &       &       &       &       &       &       &       &  \\
    Malta & MLT   &       &       &       &       &       &       &       &  \\
    Netherlands & NLD   &       &       &       &       &       &       &       &  \\
    Portugal & PRT   &       &       &       &       &       &       &       &  \\
    Slovakia & SVK   &       &       &       &       &       &       &       &  \\
    Slovenia & SVN   &       &       &       &       &       &       &       &  \\
    Spain & ESP   &       &       &       &       &       &       &       &  \\
    \botrule
    \end{tabular}%
    }
  \label{table_A1}%
\end{table}%

\newpage 

\begin{table}
  \centering
  \caption{{\bf List of WIOD industries.}}
  \resizebox{18cm}{!}{
    \begin{tabular}{@{\extracolsep{4pt}}llll@{}}
    \toprule
    \textbf{Full Name} & \textbf{ISIC Rev. 3 Code} & \textbf{WIOD Code} & \textbf{3-Letter Code} \\
    \cline{1-1} \cline{2-2} \cline{3-3} \cline{4-4}
    Agriculture, Hunting, Forestry and Fishing & AtB   & c1    & Agr \\
    Mining and Quarrying & C     & c2    & Min \\
    Food, Beverages and Tobacco & 15t16 & c3    & Fod \\
    Textiles and Textile Products & 17t18 & c4    & Tex \\
    Leather, Leather and Footwear & 19    & c5    & Lth \\
    Wood and Products of Wood and Cork & 20    & c6    & Wod \\
    Pulp, Paper, Paper , Printing and Publishing & 21t22 & c7    & Pup \\
    Coke, Refined Petroleum and Nuclear Fuel & 23    & c8    & Cok \\
    Chemicals and Chemical Products & 24    & c9    & Chm \\
    Rubber and Plastics & 25    & c10   & Rub \\
    Other Non-Metallic Mineral & 26    & c11   & Omn \\
    Basic Metals and Fabricated Metal & 27t28 & c12   & Met \\
    Machinery, Nec & 29    & c13   & Mch \\
    Electrical and Optical Equipment & 30t33 & c14   & Elc \\
    Transport Equipment & 34t35 & c15   & Tpt \\
    Manufacturing, Nec; Recycling & 36t37 & c16   & Mnf \\
    Electricity, Gas and Water Supply & E     & c17   & Ele \\
    Construction & F     & c18   & Cst \\
    Sale, Maintenance and Repair of Motor Vehicles and Motorcycles; Retail Sale of Fuel & 50    & c19   & Sal \\
    Wholesale Trade and Commission Trade, Except of Motor Vehicles and Motorcycles & 51    & c20   & Whl \\
    Retail Trade, Except of Motor Vehicles and Motorcycles; Repair of Household Goods & 52    & c21   & Rtl \\
    Hotels and Restaurants & H     & c22   & Htl \\
    Inland Transport & 60    & c23   & Ldt \\
    Water Transport & 61    & c24   & Wtt \\
    Air Transport & 62    & c25   & Ait \\
    Other Supporting and Auxiliary Transport Activities; Activities of Travel Agencies & 63    & c26   & Otr \\
    Post and Telecommunications & 64    & c27   & Pst \\
    Financial Intermediation & J     & c28   & Fin \\
    Real Estate Activities & 70    & c29   & Est \\
    Renting of M\&Eq and Other Business Activities & 71t74 & c30   & Obs \\
    Public Admin and Defence; Compulsory Social Security & L     & c31   & Pub \\
    Education & M     & c32   & Edu \\
    Health and Social Work & N     & c33   & Hth \\
    Other Community, Social and Personal Services & O     & c34   & Ocm \\
    Private Households with Employed Persons & P     & c35   & Pvt \\
    \botrule
    \end{tabular}%
    }
  \label{table_A2}%
\end{table}%

\newpage

\begin{figure}[!ht]
  \begin{center}
    \subfigure[Inflation]{\label{inflation}\includegraphics[width=8cm]{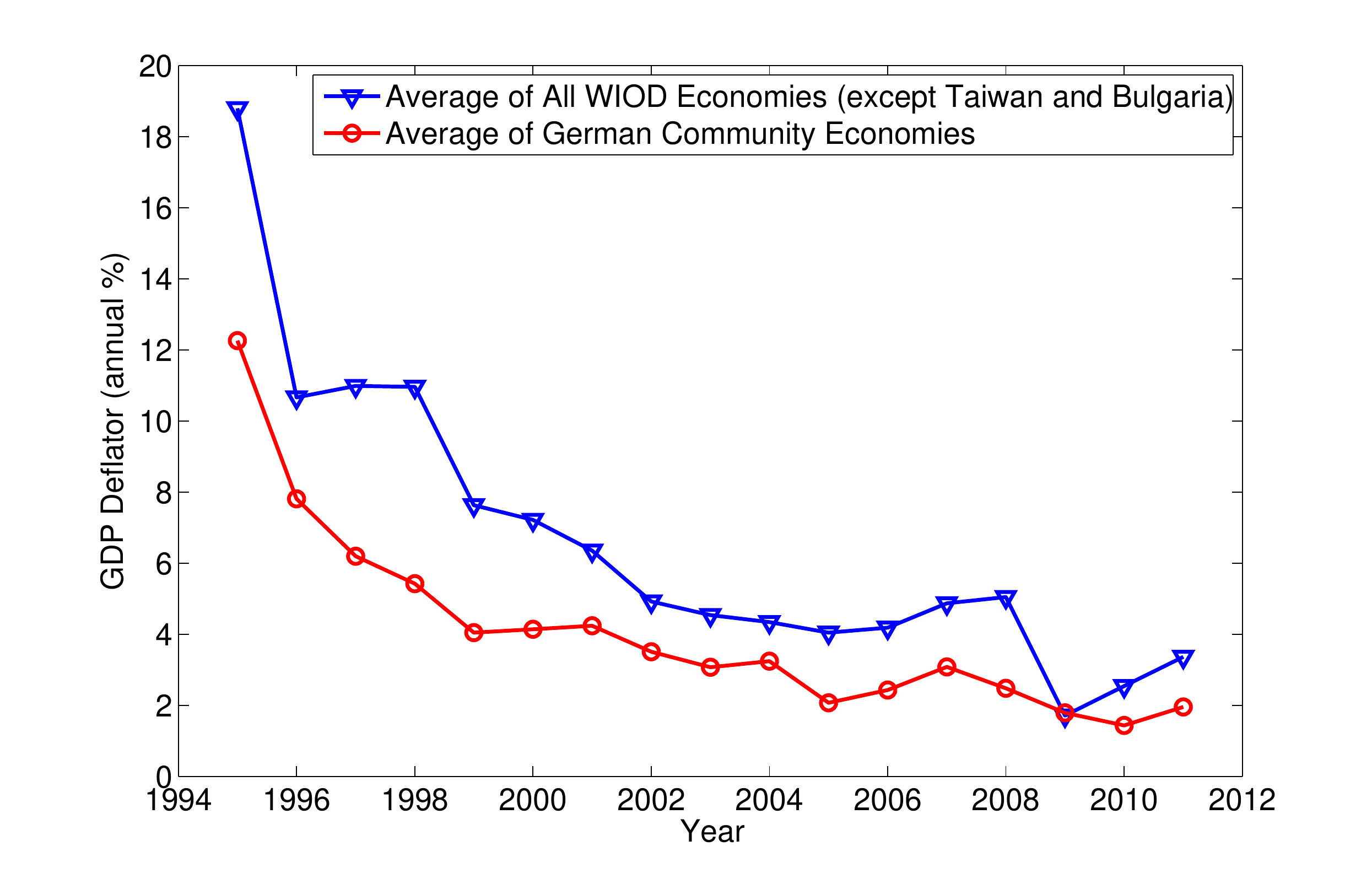}}
    \subfigure[Exchange Rate]{\label{exchange}\includegraphics[width=8cm]{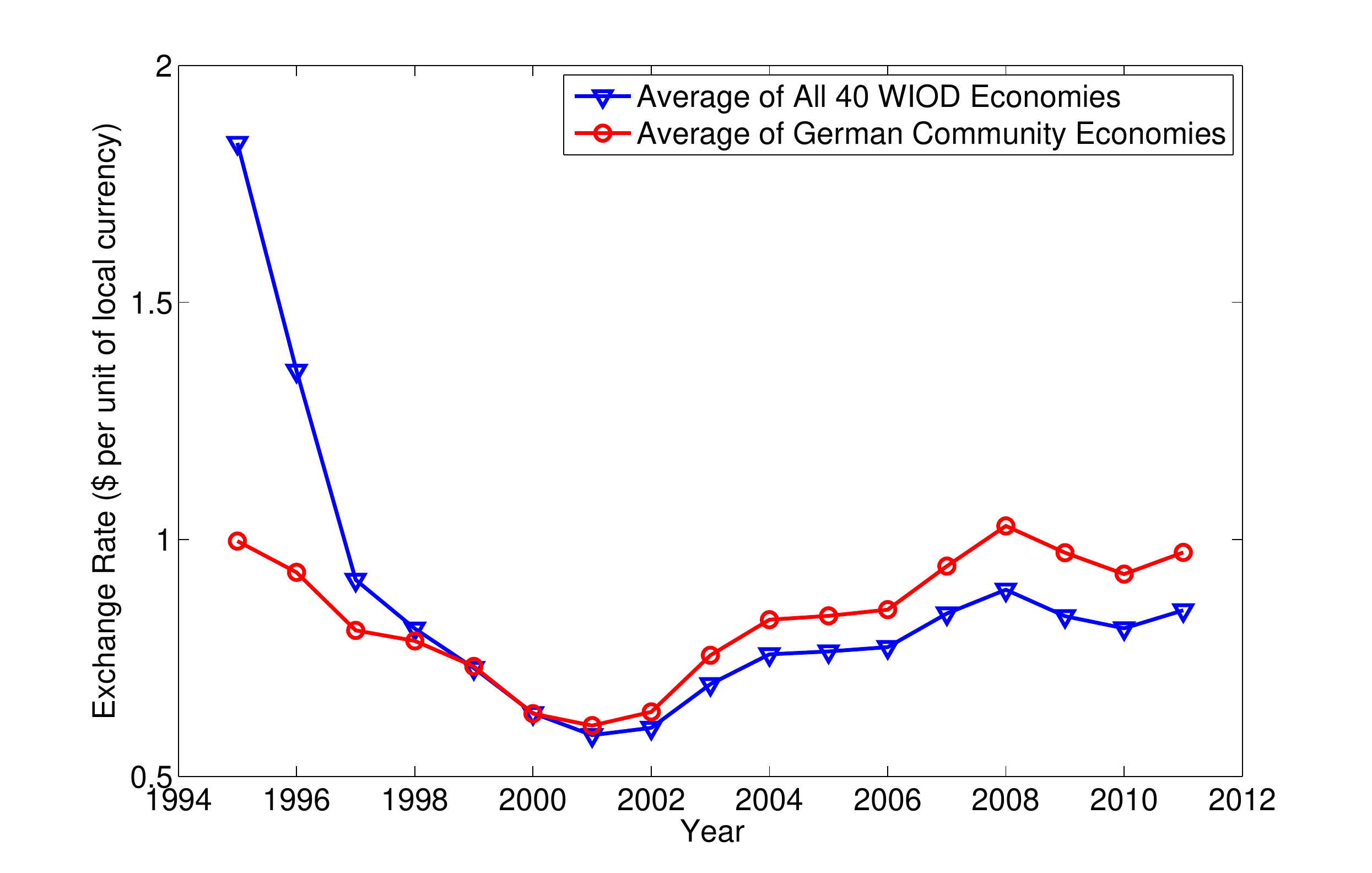}} \\
  \end{center}
  \caption{{\bf Average inflation rate and exchange rate.} (a) shows the average inflation rate of all the 40 WIOD economies (except Taiwan and Bulgaria) versus the average inflation rate of the German community. We compare the average inflation rate, i.e., the annual GDP deflator, across all the WIOD economies (except Taiwan and Bulgaria) with the average annual GDP deflator across the 9 major member economies in the German community detected in 2011, i.e., Germany, Austria, Belgium, Luxembourg, Hungary, Czech Republic, Slovakia, Slovenia, and Poland. During 1995-2011, the average inflation of the German community was almost always below that of all the WIOD economies. The data source is the World Development Indicators, the World Bank, \url{http://data.worldbank.org/indicator/NY.GDP.DEFL.KD.ZG}. (b) shows the average exchange rate of all the 40 WIOD economies versus the average exchange rate of the German community. We compare the average exchange rate, i.e., US dollars per unit of local currency, across all the WIOD economies with the average exchange rate across the 9 major economies in the German community detected in 2011, i.e., Germany, Austria, Belgium, Luxembourg, Hungary, Czech Republic, Slovakia, Slovenia, and Poland. The average exchange rate of the German community was basically below that of all the WIOD economies before 2000. Only from 2001, the community average became slightly (no more than 16\%) higher than the overall average. The data source is the exchange rate data used in the WIOD, \url{http://www.wiod.org/protected3/data/update_sep12/EXR_WIOD_Sep12.xlsx}.}
  \label{inflation_exchange}
\end{figure}

\newpage

\begin{table}
  \centering
  \caption{{\bf The most important economies by industry over time: using the final-demand-weighted backward linkage measure.}}
  \resizebox{18cm}{!}{
    \begin{tabular}{clllllllllllllllll}
    \toprule
    \textbf{Industry/Year} & \textbf{1995} & \textbf{1996} & \textbf{1997} & \textbf{1998} & \textbf{1999} & \textbf{2000} & \textbf{2001} & \textbf{2002} & \textbf{2003} & \textbf{2004} & \textbf{2005} & \textbf{2006} & \textbf{2007} & \textbf{2008} & \textbf{2009} & \textbf{2010} & \textbf{2011} \\
    \colrule
    \textbf{Agr} & CHN   & CHN   & CHN   & CHN   & CHN   & CHN   & CHN   & CHN   & CHN   & CHN   & CHN   & CHN   & CHN   & CHN   & CHN   & CHN   & CHN \\
    \textbf{Min} & USA   & USA   & USA   & USA   & USA   & USA   & USA   & USA   & USA   & USA   & USA   & USA   & USA   & USA   & USA   & USA   & USA \\
    \textbf{Fod} & USA   & USA   & USA   & USA   & USA   & USA   & USA   & USA   & USA   & USA   & USA   & USA   & USA   & USA   & USA   & USA   & CHN \\
    \textbf{Tex} & USA   & CHN   & CHN   & CHN   & CHN   & CHN   & CHN   & CHN   & CHN   & CHN   & CHN   & CHN   & CHN   & CHN   & CHN   & CHN   & CHN \\
    \textbf{Lth} & CHN   & CHN   & CHN   & CHN   & CHN   & CHN   & CHN   & CHN   & CHN   & CHN   & CHN   & CHN   & CHN   & CHN   & CHN   & CHN   & CHN \\
    \textbf{Wod} & JPN   & USA   & USA   & USA   & USA   & USA   & USA   & USA   & USA   & USA   & USA   & USA   & USA   & CHN   & CHN   & CHN   & CHN \\
    \textbf{Pup} & USA   & USA   & USA   & USA   & USA   & USA   & USA   & USA   & USA   & USA   & USA   & USA   & USA   & USA   & USA   & USA   & USA \\
    \textbf{Cok} & USA   & USA   & USA   & USA   & USA   & USA   & USA   & USA   & USA   & USA   & USA   & USA   & USA   & USA   & USA   & USA   & USA \\
    \textbf{Chm} & USA   & USA   & USA   & USA   & USA   & USA   & USA   & USA   & USA   & USA   & USA   & USA   & USA   & USA   & USA   & USA   & USA \\
    \textbf{Rub} & USA   & USA   & USA   & USA   & USA   & USA   & USA   & USA   & USA   & USA   & USA   & USA   & CHN   & CHN   & CHN   & CHN   & CHN \\
    \textbf{Omn} & CHN   & CHN   & CHN   & CHN   & CHN   & CHN   & CHN   & CHN   & CHN   & CHN   & USA   & USA   & DEU   & CHN   & CHN   & CHN   & CHN \\
    \textbf{Met} & JPN   & JPN   & JPN   & JPN   & JPN   & JPN   & JPN   & JPN   & JPN   & JPN   & JPN   & JPN   & JPN   & JPN   & CHN   & CHN   & CHN \\
    \textbf{Mch} & JPN   & JPN   & USA   & USA   & USA   & USA   & USA   & USA   & USA   & USA   & CHN   & CHN   & CHN   & CHN   & CHN   & CHN   & CHN \\
    \textbf{Elc} & JPN   & USA   & USA   & USA   & USA   & USA   & USA   & USA   & CHN   & CHN   & CHN   & CHN   & CHN   & CHN   & CHN   & CHN   & CHN \\
    \textbf{Tpt} & USA   & USA   & USA   & USA   & USA   & USA   & USA   & USA   & USA   & USA   & USA   & USA   & USA   & USA   & CHN   & CHN   & CHN \\
    \textbf{Mnf} & USA   & USA   & USA   & USA   & USA   & USA   & USA   & USA   & USA   & USA   & USA   & USA   & USA   & USA   & USA   & USA   & USA \\
    \textbf{Ele} & USA   & USA   & USA   & USA   & USA   & USA   & USA   & USA   & USA   & USA   & USA   & USA   & USA   & USA   & USA   & USA   & USA \\
    \textbf{Cst} & JPN   & JPN   & USA   & USA   & USA   & USA   & USA   & USA   & USA   & USA   & USA   & USA   & CHN   & CHN   & CHN   & CHN   & CHN \\
    \textbf{Sal} & USA   & USA   & USA   & USA   & USA   & USA   & USA   & USA   & USA   & USA   & USA   & USA   & USA   & USA   & USA   & USA   & USA \\
    \textbf{Whl} & JPN   & JPN   & USA   & USA   & USA   & USA   & USA   & USA   & USA   & USA   & USA   & USA   & USA   & USA   & USA   & USA   & USA \\
    \textbf{Rtl} & USA   & USA   & USA   & USA   & USA   & USA   & USA   & USA   & USA   & USA   & USA   & USA   & USA   & USA   & USA   & USA   & USA \\
    \textbf{Htl} & USA   & USA   & USA   & USA   & USA   & USA   & USA   & USA   & USA   & USA   & USA   & USA   & USA   & USA   & USA   & USA   & USA \\
    \textbf{Ldt} & JPN   & JPN   & JPN   & USA   & USA   & USA   & USA   & USA   & USA   & USA   & USA   & USA   & USA   & USA   & USA   & USA   & IND \\
    \textbf{Wtt} & USA   & USA   & USA   & USA   & USA   & USA   & USA   & USA   & USA   & USA   & USA   & USA   & USA   & USA   & USA   & USA   & USA \\
    \textbf{Ait} & USA   & USA   & USA   & USA   & USA   & USA   & USA   & USA   & USA   & USA   & USA   & USA   & USA   & USA   & USA   & USA   & USA \\
    \textbf{Otr} & DEU   & DEU   & DEU   & DEU   & DEU   & USA   & USA   & ITA   & ITA   & ITA   & ITA   & ITA   & ITA   & ITA   & ITA   & CHN   & CHN \\
    \textbf{Pst} & USA   & USA   & USA   & USA   & USA   & USA   & USA   & USA   & USA   & USA   & USA   & USA   & USA   & USA   & USA   & USA   & USA \\
    \textbf{Fin} & USA   & USA   & USA   & USA   & USA   & USA   & USA   & USA   & USA   & USA   & USA   & USA   & USA   & USA   & USA   & USA   & USA \\
    \textbf{Est} & USA   & USA   & USA   & USA   & USA   & USA   & USA   & USA   & USA   & USA   & USA   & USA   & USA   & USA   & USA   & USA   & USA \\
    \textbf{Obs} & USA   & USA   & USA   & USA   & USA   & USA   & USA   & USA   & USA   & USA   & USA   & USA   & USA   & USA   & USA   & USA   & USA \\
    \textbf{Pub} & USA   & USA   & USA   & USA   & USA   & USA   & USA   & USA   & USA   & USA   & USA   & USA   & USA   & USA   & USA   & USA   & USA \\
    \textbf{Edu} & JPN   & JPN   & JPN   & JPN   & JPN   & JPN   & JPN   & JPN   & JPN   & JPN   & JPN   & USA   & CHN   & CHN   & CHN   & CHN   & CHN \\
    \textbf{Hth} & USA   & USA   & USA   & USA   & USA   & USA   & USA   & USA   & USA   & USA   & USA   & USA   & USA   & USA   & USA   & USA   & USA \\
    \textbf{Ocm} & USA   & USA   & USA   & USA   & USA   & USA   & USA   & USA   & USA   & USA   & USA   & USA   & USA   & USA   & USA   & USA   & USA \\
    \textbf{Pvt} & ITA   & ITA   & ITA   & USA   & USA   & USA   & USA   & USA   & ITA   & ITA   & ITA   & ITA   & ITA   & ITA   & ITA   & ITA   & ITA \\
    \botrule
    \end{tabular}%
    }
  \label{table_A3}%
\end{table}%

\newpage

\begin{table}
  \centering
  \caption{{\bf The most important industries by economy over time: using the final-demand-weighted backward linkage measure.}}
  \resizebox{18cm}{!}{
    \begin{tabular}{clllllllllllllllll}
    \toprule
    \textbf{Economy/Year} & \textbf{1995} & \textbf{1996} & \textbf{1997} & \textbf{1998} & \textbf{1999} & \textbf{2000} & \textbf{2001} & \textbf{2002} & \textbf{2003} & \textbf{2004} & \textbf{2005} & \textbf{2006} & \textbf{2007} & \textbf{2008} & \textbf{2009} & \textbf{2010} & \textbf{2011} \\
    \colrule
    \textbf{AUS} & Cst   & Cst   & Cst   & Cst   & Cst   & Cst   & Cst   & Cst   & Cst   & Cst   & Cst   & Cst   & Cst   & Cst   & Cst   & Cst   & Cst \\
    \textbf{AUT} & Cst   & Cst   & Cst   & Cst   & Cst   & Cst   & Cst   & Cst   & Cst   & Cst   & Cst   & Cst   & Cst   & Cst   & Cst   & Cst   & Cst \\
    \textbf{BEL} & Fod   & Fod   & Cst   & Cst   & Cst   & Cst   & Cst   & Cst   & Cst   & Cst   & Cst   & Cst   & Cst   & Cst   & Cst   & Cst   & Cst \\
    \textbf{BGR} & Fod   & Fod   & Fod   & Fod   & Agr   & Agr   & Fod   & Fod   & Fod   & Fod   & Cst   & Cst   & Cst   & Cst   & Cst   & Cst   & Cst \\
    \textbf{BRA} & Pub   & Pub   & Pub   & Pub   & Pub   & Pub   & Pub   & Pub   & Pub   & Pub   & Pub   & Pub   & Pub   & Pub   & Pub   & Pub   & Pub \\
    \textbf{CAN} & Pub   & Pub   & Cst   & Pub   & Tpt   & Tpt   & Pub   & Pub   & Cst   & Cst   & Cst   & Cst   & Cst   & Cst   & Cst   & Cst   & Cst \\
    \textbf{CHN} & Cst   & Cst   & Cst   & Cst   & Cst   & Cst   & Cst   & Cst   & Cst   & Cst   & Cst   & Cst   & Cst   & Cst   & Cst   & Cst   & Cst \\
    \textbf{CYP} & Htl   & Cst   & Htl   & Htl   & Htl   & Htl   & Htl   & Pub   & Pub   & Cst   & Cst   & Cst   & Cst   & Cst   & Cst   & Cst   & Cst \\
    \textbf{CZE} & Cst   & Cst   & Cst   & Cst   & Cst   & Cst   & Cst   & Cst   & Cst   & Cst   & Cst   & Cst   & Cst   & Cst   & Cst   & Tpt   & Tpt \\
    \textbf{DEU} & Cst   & Cst   & Cst   & Cst   & Cst   & Cst   & Tpt   & Tpt   & Tpt   & Tpt   & Tpt   & Tpt   & Tpt   & Tpt   & Tpt   & Tpt   & Tpt \\
    \textbf{DNK} & Fod   & Fod   & Cst   & Cst   & Cst   & Cst   & Cst   & Hth   & Hth   & Hth   & Cst   & Cst   & Cst   & Hth   & Hth   & Hth   & Hth \\
    \textbf{ESP} & Cst   & Cst   & Cst   & Cst   & Cst   & Cst   & Cst   & Cst   & Cst   & Cst   & Cst   & Cst   & Cst   & Cst   & Cst   & Cst   & Cst \\
    \textbf{EST} & Fod   & Fod   & Fod   & Cst   & Cst   & Cst   & Cst   & Cst   & Cst   & Cst   & Cst   & Cst   & Cst   & Cst   & Cst   & Cst   & Cst \\
    \textbf{FIN} & Cst   & Cst   & Cst   & Cst   & Cst   & Cst   & Cst   & Cst   & Cst   & Cst   & Cst   & Cst   & Cst   & Cst   & Cst   & Cst   & Cst \\
    \textbf{FRA} & Cst   & Cst   & Cst   & Cst   & Cst   & Cst   & Cst   & Cst   & Cst   & Cst   & Cst   & Cst   & Cst   & Cst   & Cst   & Cst   & Cst \\
    \textbf{GBR} & Pub   & Cst   & Cst   & Cst   & Cst   & Cst   & Cst   & Cst   & Cst   & Cst   & Hth   & Hth   & Hth   & Hth   & Hth   & Hth   & Hth \\
    \textbf{GRC} & Cst   & Cst   & Cst   & Cst   & Cst   & Cst   & Cst   & Cst   & Cst   & Cst   & Cst   & Cst   & Cst   & Cst   & Pub   & Pub   & Pub \\
    \textbf{HUN} & Fod   & Fod   & Fod   & Fod   & Elc   & Elc   & Elc   & Elc   & Elc   & Elc   & Elc   & Elc   & Elc   & Elc   & Elc   & Elc   & Elc \\
    \textbf{IDN} & Cst   & Cst   & Cst   & Cst   & Fod   & Cst   & Cst   & Cst   & Cst   & Cst   & Cst   & Cst   & Cst   & Cst   & Cst   & Cst   & Cst \\
    \textbf{IND} & Agr   & Agr   & Cst   & Cst   & Cst   & Cst   & Cst   & Cst   & Cst   & Cst   & Cst   & Cst   & Cst   & Cst   & Cst   & Cst   & Cst \\
    \textbf{IRL} & Fod   & Fod   & Fod   & Elc   & Elc   & Elc   & Elc   & Elc   & Cst   & Cst   & Cst   & Cst   & Cst   & Cst   & Cst   & Chm   & Chm \\
    \textbf{ITA} & Cst   & Cst   & Cst   & Cst   & Cst   & Cst   & Cst   & Cst   & Cst   & Cst   & Cst   & Cst   & Cst   & Cst   & Cst   & Cst   & Cst \\
    \textbf{JPN} & Cst   & Cst   & Cst   & Cst   & Cst   & Cst   & Cst   & Cst   & Cst   & Cst   & Cst   & Cst   & Cst   & Cst   & Cst   & Cst   & Cst \\
    \textbf{KOR} & Cst   & Cst   & Cst   & Cst   & Cst   & Cst   & Cst   & Cst   & Cst   & Cst   & Cst   & Cst   & Cst   & Cst   & Cst   & Cst   & Cst \\
    \textbf{LTU} & Fod   & Fod   & Pub   & Cst   & Cst   & Fod   & Fod   & Cst   & Cst   & Cst   & Cst   & Cst   & Cst   & Cst   & Fod   & Fod   & Fod \\
    \textbf{LUX} & Cst   & Fin   & Fin   & Fin   & Fin   & Fin   & Fin   & Fin   & Fin   & Fin   & Fin   & Fin   & Fin   & Fin   & Fin   & Fin   & Fin \\
    \textbf{LVA} & Fod   & Fod   & Fod   & Cst   & Cst   & Cst   & Cst   & Cst   & Cst   & Cst   & Cst   & Cst   & Cst   & Cst   & Cst   & Cst   & Cst \\
    \textbf{MEX} & Fod   & Fod   & Fod   & Fod   & Fod   & Cst   & Cst   & Cst   & Cst   & Cst   & Cst   & Cst   & Cst   & Cst   & Cst   & Cst   & Cst \\
    \textbf{MLT} & Elc   & Elc   & Elc   & Elc   & Elc   & Elc   & Elc   & Cst   & Cst   & Elc   & Cst   & Ocm   & Ocm   & Ocm   & Ocm   & Ocm   & Ocm \\
    \textbf{NLD} & Fod   & Fod   & Fod   & Fod   & Cst   & Cst   & Cst   & Cst   & Pub   & Pub   & Pub   & Pub   & Cst   & Cst   & Pub   & Pub   & Pub \\
    \textbf{POL} & Fod   & Fod   & Fod   & Cst   & Cst   & Cst   & Cst   & Fod   & Fod   & Fod   & Fod   & Fod   & Cst   & Cst   & Cst   & Cst   & Cst \\
    \textbf{PRT} & Cst   & Cst   & Cst   & Cst   & Cst   & Cst   & Cst   & Cst   & Cst   & Cst   & Cst   & Cst   & Cst   & Cst   & Cst   & Cst   & Cst \\
    \textbf{ROM} & Fod   & Fod   & Fod   & Fod   & Fod   & Fod   & Fod   & Fod   & Fod   & Fod   & Cst   & Cst   & Cst   & Cst   & Cst   & Cst   & Cst \\
    \textbf{RUS} & Fod   & Cst   & Cst   & Cst   & Fod   & Fod   & Cst   & Cst   & Cst   & Cst   & Cst   & Cst   & Cst   & Cst   & Cst   & Cst   & Cst \\
    \textbf{SVK} & Cst   & Cst   & Cst   & Cst   & Cst   & Cst   & Tpt   & Tpt   & Tpt   & Tpt   & Cst   & Tpt   & Tpt   & Tpt   & Cst   & Cst   & Cst \\
    \textbf{SVN} & Cst   & Cst   & Cst   & Cst   & Cst   & Cst   & Cst   & Cst   & Cst   & Cst   & Cst   & Cst   & Cst   & Cst   & Cst   & Cst   & Cst \\
    \textbf{SWE} & Est   & Est   & Est   & Est   & Est   & Hth   & Hth   & Hth   & Hth   & Hth   & Hth   & Hth   & Hth   & Hth   & Hth   & Hth   & Hth \\
    \textbf{TUR} & Cst   & Cst   & Cst   & Cst   & Fod   & Fod   & Fod   & Tex   & Tex   & Tex   & Fod   & Fod   & Tex   & Fod   & Fod   & Fod   & Fod \\
    \textbf{TWN} & Cst   & Cst   & Pub   & Pub   & Elc   & Elc   & Elc   & Elc   & Elc   & Cst   & Cst   & Cst   & Cst   & Cst   & Cst   & Cst   & Cst \\
    \textbf{USA} & Pub   & Pub   & Pub   & Pub   & Pub   & Pub   & Pub   & Pub   & Pub   & Pub   & Pub   & Pub   & Pub   & Pub   & Pub   & Pub   & Pub \\
    \botrule
    \end{tabular}%
    }
  \label{table_A4}%
\end{table}%

\newpage

\begin{table}
  \centering
  \caption{{\bf The most important economies by industry over time: using the PageRank centrality measure.}}
  \resizebox{18cm}{!}{
    \begin{tabular}{clllllllllllllllll}
    \toprule
    \textbf{Industry/Year} & \textbf{1995} & \textbf{1996} & \textbf{1997} & \textbf{1998} & \textbf{1999} & \textbf{2000} & \textbf{2001} & \textbf{2002} & \textbf{2003} & \textbf{2004} & \textbf{2005} & \textbf{2006} & \textbf{2007} & \textbf{2008} & \textbf{2009} & \textbf{2010} & \textbf{2011} \\
    \colrule
    \textbf{Agr} & RUS   & RUS   & RUS   & RUS   & DEU   & BGR   & BGR   & CHN   & CHN   & CHN   & CHN   & CHN   & CHN   & RUS   & CHN   & CHN   & CHN \\
    \textbf{Min} & RUS   & RUS   & RUS   & RUS   & RUS   & RUS   & RUS   & RUS   & RUS   & RUS   & RUS   & RUS   & RUS   & RUS   & RUS   & RUS   & RUS \\
    \textbf{Fod} & DEU   & DEU   & DEU   & USA   & DEU   & USA   & USA   & USA   & USA   & USA   & USA   & USA   & USA   & USA   & USA   & USA   & USA \\
    \textbf{Tex} & ITA   & ITA   & ITA   & ITA   & ITA   & TUR   & ITA   & TUR   & TUR   & TUR   & TUR   & TUR   & TUR   & TUR   & TUR   & TUR   & CHN \\
    \textbf{Lth} & ITA   & ITA   & ITA   & ITA   & ITA   & ITA   & ITA   & CHN   & CHN   & CHN   & ITA   & ITA   & ITA   & ITA   & CHN   & CHN   & CHN \\
    \textbf{Wod} & DEU   & DEU   & USA   & USA   & USA   & USA   & USA   & USA   & LVA   & USA   & USA   & USA   & LVA   & CHN   & CHN   & CHN   & CHN \\
    \textbf{Pup} & USA   & USA   & USA   & USA   & USA   & USA   & USA   & USA   & USA   & USA   & USA   & USA   & USA   & USA   & USA   & USA   & USA \\
    \textbf{Cok} & BRA   & USA   & BRA   & BRA   & BRA   & USA   & USA   & USA   & DEU   & DEU   & USA   & USA   & USA   & RUS   & RUS   & USA   & FRA \\
    \textbf{Chm} & DEU   & USA   & USA   & USA   & USA   & USA   & USA   & USA   & USA   & USA   & USA   & USA   & USA   & USA   & USA   & USA   & CHN \\
    \textbf{Rub} & DEU   & DEU   & USA   & DEU   & USA   & USA   & USA   & USA   & USA   & DEU   & DEU   & DEU   & DEU   & CHN   & CHN   & CHN   & CHN \\
    \textbf{Omn} & CHN   & CHN   & CHN   & CHN   & CHN   & CHN   & CHN   & CHN   & CHN   & CHN   & CHN   & CHN   & CHN   & CHN   & CHN   & CHN   & CHN \\
    \textbf{Met} & DEU   & DEU   & DEU   & DEU   & USA   & DEU   & DEU   & DEU   & DEU   & DEU   & DEU   & CHN   & CHN   & CHN   & CHN   & CHN   & CHN \\
    \textbf{Mch} & DEU   & DEU   & DEU   & DEU   & DEU   & DEU   & DEU   & DEU   & DEU   & DEU   & DEU   & DEU   & DEU   & DEU   & DEU   & DEU   & DEU \\
    \textbf{Elc} & USA   & USA   & USA   & USA   & USA   & USA   & DEU   & DEU   & CHN   & CHN   & CHN   & CHN   & CHN   & CHN   & CHN   & CHN   & CHN \\
    \textbf{Tpt} & USA   & DEU   & USA   & DEU   & DEU   & DEU   & DEU   & DEU   & DEU   & DEU   & DEU   & DEU   & DEU   & DEU   & DEU   & DEU   & DEU \\
    \textbf{Mnf} & DEU   & DEU   & ITA   & DEU   & ITA   & DEU   & DEU   & ITA   & ITA   & DEU   & DEU   & DEU   & DEU   & DEU   & DEU   & DEU   & DEU \\
    \textbf{Ele} & FRA   & FRA   & RUS   & FRA   & DEU   & USA   & USA   & DEU   & DEU   & DEU   & DEU   & DEU   & DEU   & DEU   & DEU   & RUS   & RUS \\
    \textbf{Cst} & DEU   & DEU   & DEU   & USA   & USA   & USA   & USA   & ESP   & ESP   & ESP   & ESP   & ESP   & ESP   & ESP   & ESP   & CHN   & CHN \\
    \textbf{Sal} & ROM   & ROM   & ROM   & ROM   & ROM   & ROM   & ROM   & ROM   & ITA   & ITA   & ITA   & ITA   & ITA   & ITA   & ITA   & ITA   & ITA \\
    \textbf{Whl} & ITA   & ITA   & ITA   & ITA   & ITA   & ITA   & ITA   & ITA   & ITA   & ITA   & ITA   & ITA   & RUS   & RUS   & RUS   & RUS   & RUS \\
    \textbf{Rtl} & USA   & USA   & USA   & USA   & USA   & USA   & GBR   & GBR   & GBR   & GBR   & USA   & USA   & GBR   & USA   & GBR   & USA   & USA \\
    \textbf{Htl} & USA   & USA   & USA   & USA   & USA   & USA   & USA   & USA   & USA   & USA   & USA   & USA   & USA   & USA   & USA   & USA   & USA \\
    \textbf{Ldt} & IND   & IND   & IND   & IND   & IND   & IND   & IND   & IND   & IND   & IND   & IND   & IND   & IND   & IND   & IND   & IND   & IND \\
    \textbf{Wtt} & JPN   & JPN   & JPN   & JPN   & JPN   & JPN   & JPN   & DNK   & JPN   & JPN   & JPN   & JPN   & JPN   & JPN   & JPN   & JPN   & JPN \\
    \textbf{Ait} & CYP   & CYP   & CYP   & CYP   & CYP   & CYP   & CYP   & CYP   & CYP   & CYP   & CYP   & DEU   & DEU   & DEU   & DEU   & DEU   & DEU \\
    \textbf{Otr} & DEU   & DEU   & DEU   & DEU   & DEU   & DEU   & DEU   & DEU   & SWE   & DEU   & DEU   & DEU   & DEU   & DEU   & DEU   & SWE   & SWE \\
    \textbf{Pst} & USA   & USA   & USA   & USA   & USA   & KOR   & USA   & USA   & USA   & USA   & USA   & USA   & USA   & USA   & USA   & USA   & USA \\
    \textbf{Fin} & USA   & USA   & USA   & USA   & USA   & USA   & USA   & USA   & USA   & USA   & USA   & USA   & USA   & USA   & USA   & USA   & USA \\
    \textbf{Est} & USA   & USA   & USA   & USA   & USA   & USA   & USA   & USA   & USA   & USA   & USA   & USA   & USA   & USA   & USA   & USA   & USA \\
    \textbf{Obs} & USA   & USA   & USA   & USA   & USA   & USA   & USA   & USA   & USA   & USA   & USA   & USA   & USA   & USA   & USA   & USA   & USA \\
    \textbf{Pub} & USA   & USA   & USA   & USA   & USA   & USA   & USA   & USA   & USA   & USA   & USA   & USA   & USA   & USA   & USA   & USA   & USA \\
    \textbf{Edu} & RUS   & RUS   & RUS   & RUS   & GBR   & GBR   & GBR   & GBR   & CHN   & GBR   & CHN   & CHN   & CHN   & CHN   & DEU   & CHN   & CHN \\
    \textbf{Hth} & USA   & USA   & USA   & USA   & USA   & USA   & USA   & USA   & USA   & USA   & GBR   & GBR   & GBR   & GBR   & GBR   & GBR   & GBR \\
    \textbf{Ocm} & USA   & USA   & USA   & USA   & USA   & USA   & USA   & USA   & USA   & USA   & USA   & USA   & USA   & USA   & USA   & USA   & USA \\
    \textbf{Pvt} & IND   & IND   & IND   & IND   & IND   & IND   & IND   & IND   & IND   & IND   & IND   & IND   & IND   & IND   & IND   & IND   & IND \\
    \botrule
    \end{tabular}%
    }
  \label{table_A5}%
\end{table}%

\newpage

\begin{table}
  \centering
  \caption{{\bf The most important industries by economy over time: using the PageRank centrality measure.}}
  \resizebox{18cm}{!}{
    \begin{tabular}{clllllllllllllllll}
    \toprule
    \textbf{Economy/Year} & \textbf{1995} & \textbf{1996} & \textbf{1997} & \textbf{1998} & \textbf{1999} & \textbf{2000} & \textbf{2001} & \textbf{2002} & \textbf{2003} & \textbf{2004} & \textbf{2005} & \textbf{2006} & \textbf{2007} & \textbf{2008} & \textbf{2009} & \textbf{2010} & \textbf{2011} \\
    \colrule
    \textbf{AUS} & Cst   & Cst   & Cst   & Cst   & Cst   & Cst   & Cst   & Cst   & Cst   & Cst   & Cst   & Cst   & Cst   & Cst   & Cst   & Cst   & Cst \\
    \textbf{AUT} & Cst   & Cst   & Cst   & Cst   & Cst   & Cst   & Cst   & Cst   & Hth   & Cst   & Cst   & Cst   & Cst   & Cst   & Cst   & Cst   & Cst \\
    \textbf{BEL} & Hth   & Hth   & Hth   & Hth   & Cst   & Hth   & Cst   & Hth   & Hth   & Hth   & Cst   & Cst   & Cst   & Cst   & Cst   & Cst   & Cst \\
    \textbf{BGR} & Fod   & Fod   & Agr   & Agr   & Agr   & Agr   & Agr   & Agr   & Agr   & Agr   & Agr   & Agr   & Fod   & Cst   & Cst   & Cst   & Cst \\
    \textbf{BRA} & Fod   & Fod   & Fod   & Fod   & Fod   & Fod   & Fod   & Fod   & Fod   & Fod   & Fod   & Fod   & Tpt   & Tpt   & Fod   & Tpt   & Tpt \\
    \textbf{CAN} & Pub   & Pub   & Pub   & Pub   & Pub   & Pub   & Pub   & Pub   & Pub   & Pub   & Pub   & Pub   & Pub   & Pub   & Pub   & Tpt   & Tpt \\
    \textbf{CHN} & Tex   & Tex   & Met   & Tex   & Tex   & Elc   & Tex   & Elc   & Elc   & Elc   & Elc   & Elc   & Elc   & Elc   & Elc   & Elc   & Elc \\
    \textbf{CYP} & Fod   & Fod   & Fod   & Pub   & Fod   & Fod   & Pub   & Fod   & Fod   & Fod   & Pub   & Pub   & Pub   & Cst   & Pub   & Pub   & Pub \\
    \textbf{CZE} & Cst   & Cst   & Cst   & Cst   & Cst   & Cst   & Cst   & Cst   & Cst   & Cst   & Cst   & Cst   & Cst   & Cst   & Cst   & Tpt   & Tpt \\
    \textbf{DEU} & Tpt   & Tpt   & Tpt   & Tpt   & Tpt   & Tpt   & Tpt   & Tpt   & Tpt   & Tpt   & Tpt   & Tpt   & Tpt   & Tpt   & Tpt   & Tpt   & Tpt \\
    \textbf{DNK} & Fod   & Fod   & Fod   & Fod   & Fod   & Fod   & Fod   & Fod   & Fod   & Fod   & Fod   & Obs   & Obs   & Obs   & Obs   & Obs   & Obs \\
    \textbf{ESP} & Cst   & Cst   & Cst   & Cst   & Cst   & Cst   & Cst   & Cst   & Cst   & Cst   & Cst   & Cst   & Cst   & Cst   & Cst   & Cst   & Cst \\
    \textbf{EST} & Fod   & Fod   & Fod   & Fod   & Otr   & Otr   & Otr   & Otr   & Otr   & Otr   & Otr   & Cst   & Cst   & Cst   & Otr   & Otr   & Otr \\
    \textbf{FIN} & Pup   & Pup   & Pup   & Cst   & Cst   & Elc   & Cst   & Cst   & Cst   & Cst   & Cst   & Hth   & Hth   & Hth   & Hth   & Hth   & Hth \\
    \textbf{FRA} & Tpt   & Tpt   & Tpt   & Tpt   & Tpt   & Tpt   & Tpt   & Tpt   & Tpt   & Tpt   & Tpt   & Tpt   & Tpt   & Tpt   & Tpt   & Tpt   & Cst \\
    \textbf{GBR} & Cst   & Cst   & Cst   & Cst   & Hth   & Hth   & Hth   & Hth   & Hth   & Hth   & Hth   & Hth   & Hth   & Hth   & Hth   & Hth   & Hth \\
    \textbf{GRC} & Pub   & Fod   & Fod   & Fod   & Pub   & Pub   & Pub   & Fod   & Cst   & Fod   & Fod   & Cst   & Ocm   & Fod   & Ocm   & Ocm   & Ocm \\
    \textbf{HUN} & Fod   & Fod   & Fod   & Fod   & Fod   & Fod   & Fod   & Fod   & Fod   & Fod   & Fod   & Fod   & Fod   & Fod   & Fod   & Fod   & Agr \\
    \textbf{IDN} & Cst   & Cst   & Tex   & Fod   & Fod   & Fod   & Fod   & Fod   & Fod   & Cst   & Cst   & Cst   & Cst   & Cst   & Cst   & Cst   & Cst \\
    \textbf{IND} & Fod   & Ldt   & Ldt   & Ldt   & Ldt   & Fod   & Ldt   & Fod   & Fod   & Cst   & Cst   & Cst   & Cst   & Cst   & Cst   & Cst   & Cst \\
    \textbf{IRL} & Cst   & Cst   & Cst   & Cst   & Cst   & Cst   & Cst   & Cst   & Cst   & Cst   & Cst   & Cst   & Cst   & Cst   & Hth   & Hth   & Hth \\
    \textbf{ITA} & Cst   & Cst   & Cst   & Tex   & Tex   & Tex   & Tex   & Hth   & Hth   & Hth   & Hth   & Hth   & Hth   & Hth   & Hth   & Hth   & Hth \\
    \textbf{JPN} & Cst   & Cst   & Tpt   & Cst   & Cst   & Tpt   & Tpt   & Tpt   & Tpt   & Tpt   & Tpt   & Tpt   & Tpt   & Tpt   & Tpt   & Tpt   & Tpt \\
    \textbf{KOR} & Tpt   & Tpt   & Tpt   & Tpt   & Tpt   & Tpt   & Tpt   & Tpt   & Tpt   & Tpt   & Tpt   & Tpt   & Tpt   & Tpt   & Tpt   & Tpt   & Tpt \\
    \textbf{LTU} & Cst   & Cst   & Pub   & Cst   & Cst   & Fod   & Fod   & Cst   & Cst   & Cst   & Cst   & Cst   & Cst   & Cst   & Cst   & Fod   & Fod \\
    \textbf{LUX} & Fin   & Fin   & Fin   & Fin   & Fin   & Fin   & Fin   & Fin   & Fin   & Fin   & Fin   & Fin   & Fin   & Fin   & Fin   & Fin   & Fin \\
    \textbf{LVA} & Agr   & Otr   & Fod   & Fod   & Cst   & Cst   & Cst   & Cst   & Cst   & Cst   & Cst   & Cst   & Cst   & Cst   & Cst   & Cst   & Cst \\
    \textbf{MEX} & Fod   & Fod   & Cst   & Cst   & Fod   & Cst   & Cst   & Cst   & Cst   & Cst   & Cst   & Cst   & Cst   & Cst   & Cst   & Cst   & Cst \\
    \textbf{MLT} & Fod   & Fod   & Fod   & Fod   & Fod   & Fod   & Fod   & Fod   & Fod   & Fod   & Ocm   & Ocm   & Ocm   & Ocm   & Ocm   & Ocm   & Ocm \\
    \textbf{NLD} & Fod   & Cst   & Fod   & Cst   & Cst   & Cst   & Cst   & Cst   & Cst   & Cst   & Cst   & Cst   & Cst   & Cst   & Cst   & Fod   & Fod \\
    \textbf{POL} & Fod   & Fod   & Fod   & Fod   & Fod   & Cst   & Fod   & Fod   & Fod   & Fod   & Fod   & Fod   & Fod   & Cst   & Cst   & Cst   & Cst \\
    \textbf{PRT} & Cst   & Cst   & Cst   & Cst   & Cst   & Cst   & Cst   & Cst   & Cst   & Cst   & Cst   & Cst   & Cst   & Cst   & Hth   & Hth   & Hth \\
    \textbf{ROM} & Sal   & Fod   & Sal   & Sal   & Sal   & Sal   & Sal   & Sal   & Fod   & Cst   & Fod   & Cst   & Cst   & Cst   & Cst   & Cst   & Cst \\
    \textbf{RUS} & Hth   & Hth   & Hth   & Hth   & Fod   & Hth   & Hth   & Hth   & Hth   & Hth   & Hth   & Hth   & Hth   & Hth   & Hth   & Hth   & Hth \\
    \textbf{SVK} & Pub   & Pub   & Cst   & Pub   & Pub   & Cst   & Ele   & Ele   & Ele   & Cst   & Cst   & Cst   & Cst   & Cst   & Cst   & Cst   & Cst \\
    \textbf{SVN} & Cst   & Cst   & Cst   & Cst   & Cst   & Cst   & Cst   & Cst   & Cst   & Cst   & Cst   & Cst   & Cst   & Cst   & Cst   & Cst   & Cst \\
    \textbf{SWE} & Hth   & Obs   & Obs   & Hth   & Obs   & Obs   & Obs   & Tpt   & Tpt   & Tpt   & Tpt   & Tpt   & Tpt   & Tpt   & Obs   & Tpt   & Tpt \\
    \textbf{TUR} & Tex   & Tex   & Fod   & Fod   & Tex   & Tex   & Tex   & Tex   & Tex   & Tex   & Tex   & Tex   & Tex   & Tex   & Tex   & Tex   & Tex \\
    \textbf{TWN} & Fod   & Pub   & Pub   & Pub   & Elc   & Elc   & Elc   & Elc   & Elc   & Elc   & Elc   & Elc   & Elc   & Elc   & Elc   & Elc   & Elc \\
    \textbf{USA} & Pub   & Pub   & Pub   & Pub   & Pub   & Pub   & Pub   & Hth   & Hth   & Hth   & Hth   & Hth   & Pub   & Pub   & Pub   & Pub   & Pub \\
    \botrule
    \end{tabular}%
    }
  \label{table_A6}%
\end{table}%

\newpage

\begin{table}
  \centering
  \caption{{\bf The most important economies by industry over time: using the community coreness measure.}}
  \resizebox{18cm}{!}{
    \begin{tabular}{clllllllllllllllll}
    \toprule
    \textbf{Industry/Year} & \textbf{1995} & \textbf{1996} & \textbf{1997} & \textbf{1998} & \textbf{1999} & \textbf{2000} & \textbf{2001} & \textbf{2002} & \textbf{2003} & \textbf{2004} & \textbf{2005} & \textbf{2006} & \textbf{2007} & \textbf{2008} & \textbf{2009} & \textbf{2010} & \textbf{2011} \\
    \colrule
    \textbf{Agr} & USA   & USA   & USA   & USA   & CHN   & USA   & CHN   & CHN   & CHN   & CHN   & USA   & CHN   & CHN   & CHN   & CHN   & CHN   & CHN \\
    \textbf{Min} & USA   & USA   & USA   & CHN   & CHN   & USA   & USA   & USA   & CHN   & CHN   & USA   & CHN   & CHN   & CHN   & CHN   & CHN   & CHN \\
    \textbf{Fod} & JPN   & USA   & USA   & USA   & JPN   & USA   & USA   & USA   & CHN   & CHN   & CHN   & CHN   & CHN   & CHN   & CHN   & CHN   & CHN \\
    \textbf{Tex} & CHN   & CHN   & CHN   & CHN   & CHN   & CHN   & CHN   & CHN   & CHN   & CHN   & BRA   & CHN   & CHN   & CHN   & CHN   & CHN   & CHN \\
    \textbf{Lth} & CHN   & CHN   & CHN   & CHN   & CHN   & CHN   & CHN   & CHN   & CHN   & CHN   & BRA   & CHN   & CHN   & CHN   & CHN   & CHN   & CHN \\
    \textbf{Wod} & JPN   & JPN   & JPN   & JPN   & JPN   & JPN   & JPN   & JPN   & JPN   & JPN   & CHN   & CHN   & CHN   & CHN   & CHN   & CHN   & CHN \\
    \textbf{Pup} & USA   & USA   & USA   & JPN   & JPN   & JPN   & JPN   & JPN   & JPN   & JPN   & USA   & USA   & USA   & USA   & USA   & USA   & USA \\
    \textbf{Cok} & USA   & USA   & USA   & JPN   & USA   & USA   & USA   & CHN   & USA   & CHN   & USA   & USA   & USA   & USA   & CHN   & USA   & USA \\
    \textbf{Chm} & JPN   & JPN   & JPN   & JPN   & JPN   & JPN   & JPN   & JPN   & CHN   & CHN   & CHN   & CHN   & CHN   & CHN   & CHN   & CHN   & CHN \\
    \textbf{Rub} & USA   & USA   & USA   & USA   & USA   & USA   & USA   & USA   & USA   & USA   & USA   & USA   & CHN   & CHN   & CHN   & CHN   & CHN \\
    \textbf{Omn} & CHN   & CHN   & CHN   & CHN   & CHN   & CHN   & CHN   & CHN   & CHN   & CHN   & CHN   & CHN   & CHN   & CHN   & CHN   & CHN   & CHN \\
    \textbf{Met} & JPN   & JPN   & JPN   & USA   & USA   & USA   & USA   & USA   & CHN   & CHN   & CHN   & CHN   & CHN   & CHN   & CHN   & CHN   & CHN \\
    \textbf{Mch} & USA   & USA   & USA   & USA   & USA   & USA   & USA   & CHN   & CHN   & CHN   & CHN   & CHN   & CHN   & CHN   & CHN   & CHN   & CHN \\
    \textbf{Elc} & USA   & USA   & USA   & USA   & USA   & CHN   & CHN   & CHN   & CHN   & CHN   & CHN   & CHN   & CHN   & CHN   & CHN   & CHN   & CHN \\
    \textbf{Tpt} & USA   & USA   & USA   & USA   & USA   & USA   & USA   & USA   & USA   & USA   & USA   & USA   & USA   & CHN   & CHN   & CHN   & CHN \\
    \textbf{Mnf} & USA   & USA   & USA   & USA   & USA   & USA   & USA   & USA   & USA   & USA   & USA   & USA   & USA   & USA   & USA   & USA   & USA \\
    \textbf{Ele} & JPN   & USA   & USA   & USA   & USA   & USA   & USA   & USA   & JPN   & USA   & USA   & USA   & USA   & USA   & JPN   & JPN   & JPN \\
    \textbf{Cst} & JPN   & JPN   & JPN   & JPN   & JPN   & JPN   & JPN   & CHN   & CHN   & CHN   & USA   & USA   & CHN   & CHN   & CHN   & CHN   & CHN \\
    \textbf{Sal} & JPN   & JPN   & JPN   & JPN   & JPN   & JPN   & JPN   & JPN   & JPN   & JPN   & JPN   & JPN   & JPN   & JPN   & JPN   & JPN   & JPN \\
    \textbf{Whl} & JPN   & USA   & USA   & USA   & USA   & USA   & USA   & USA   & USA   & USA   & USA   & USA   & USA   & USA   & USA   & USA   & USA \\
    \textbf{Rtl} & USA   & USA   & USA   & USA   & USA   & USA   & USA   & USA   & USA   & USA   & USA   & USA   & USA   & USA   & USA   & USA   & USA \\
    \textbf{Htl} & JPN   & JPN   & JPN   & JPN   & JPN   & JPN   & JPN   & JPN   & JPN   & JPN   & JPN   & JPN   & JPN   & JPN   & JPN   & JPN   & JPN \\
    \textbf{Ldt} & JPN   & JPN   & JPN   & JPN   & JPN   & JPN   & JPN   & JPN   & JPN   & JPN   & USA   & USA   & USA   & USA   & IND   & IND   & IND \\
    \textbf{Wtt} & USA   & USA   & CHN   & CHN   & CHN   & CHN   & CHN   & CHN   & CHN   & CHN   & BGR   & CHN   & CHN   & CHN   & CHN   & CHN   & CHN \\
    \textbf{Ait} & USA   & USA   & USA   & USA   & USA   & USA   & USA   & USA   & USA   & USA   & USA   & USA   & USA   & USA   & USA   & USA   & USA \\
    \textbf{Otr} & USA   & USA   & USA   & USA   & USA   & USA   & USA   & USA   & USA   & USA   & BEL   & USA   & USA   & USA   & USA   & USA   & USA \\
    \textbf{Pst} & USA   & USA   & USA   & USA   & USA   & USA   & USA   & USA   & USA   & USA   & USA   & USA   & USA   & USA   & USA   & USA   & USA \\
    \textbf{Fin} & JPN   & JPN   & JPN   & USA   & USA   & USA   & USA   & USA   & USA   & USA   & USA   & USA   & USA   & USA   & USA   & USA   & USA \\
    \textbf{Est} & USA   & USA   & USA   & USA   & USA   & USA   & USA   & USA   & USA   & USA   & USA   & USA   & USA   & USA   & USA   & USA   & USA \\
    \textbf{Obs} & USA   & USA   & USA   & USA   & USA   & USA   & USA   & USA   & USA   & USA   & USA   & USA   & USA   & USA   & USA   & USA   & USA \\
    \textbf{Pub} & USA   & USA   & USA   & USA   & USA   & USA   & USA   & USA   & USA   & USA   & USA   & USA   & USA   & USA   & USA   & USA   & USA \\
    \textbf{Edu} & JPN   & JPN   & JPN   & JPN   & JPN   & USA   & USA   & USA   & USA   & CHN   & CHN   & CHN   & CHN   & CHN   & CHN   & CHN   & CHN \\
    \textbf{Hth} & USA   & USA   & USA   & USA   & USA   & USA   & USA   & USA   & USA   & USA   & USA   & USA   & USA   & USA   & USA   & USA   & USA \\
    \textbf{Ocm} & JPN   & USA   & USA   & USA   & USA   & USA   & USA   & USA   & USA   & USA   & USA   & USA   & USA   & USA   & USA   & USA   & USA \\
    \textbf{Pvt} & USA   & USA   & USA   & USA   & USA   & USA   & USA   & USA   & USA   & USA   & BGR   & USA   & USA   & USA   & USA   & USA   & USA \\
    \botrule
    \end{tabular}%
    }
  \label{table_A7}%
\end{table}%

\newpage

\begin{table}
  \centering
  \caption{{\bf The most important industries by economy over time: using the community coreness measure.}}
  \resizebox{18cm}{!}{
    \begin{tabular}{clllllllllllllllll}
    \toprule
    \textbf{Economy/Year} & \textbf{1995} & \textbf{1996} & \textbf{1997} & \textbf{1998} & \textbf{1999} & \textbf{2000} & \textbf{2001} & \textbf{2002} & \textbf{2003} & \textbf{2004} & \textbf{2005} & \textbf{2006} & \textbf{2007} & \textbf{2008} & \textbf{2009} & \textbf{2010} & \textbf{2011} \\
    \colrule
    \textbf{AUS} & Obs   & Obs   & Obs   & Obs   & Obs   & Obs   & Obs   & Obs   & Obs   & Obs   & Fod   & Obs   & Obs   & Obs   & Obs   & Obs   & Obs \\
    \textbf{AUT} & Fin   & Fin   & Fin   & Fin   & Cst   & Whl   & Obs   & Est   & Est   & Met   & Pub   & Cst   & Obs   & Obs   & Met   & Met   & Met \\
    \textbf{BEL} & Obs   & Obs   & Obs   & Obs   & Obs   & Obs   & Obs   & Obs   & Obs   & Obs   & Otr   & Obs   & Obs   & Obs   & Obs   & Obs   & Chm \\
    \textbf{BGR} & Fod   & Fod   & Ele   & Ele   & Agr   & Ele   & Agr   & Agr   & Agr   & Cst   & Ocm   & Cok   & Cst   & Cst   & Cst   & Cst   & Cst \\
    \textbf{BRA} & Fod   & Fod   & Obs   & Fod   & Fod   & Fod   & Fod   & Fod   & Fod   & Fod   & Tex   & Cok   & Fod   & Fod   & Fod   & Fod   & Min \\
    \textbf{CAN} & Cst   & Cst   & Cst   & Cst   & Whl   & Obs   & Obs   & Whl   & Whl   & Obs   & Ldt   & Whl   & Obs   & Obs   & Cst   & Cst   & Cst \\
    \textbf{CHN} & Cst   & Cst   & Cst   & Cst   & Cst   & Cst   & Cst   & Cst   & Cst   & Cst   & Cst   & Met   & Met   & Met   & Cst   & Cst   & Cst \\
    \textbf{CYP} & Cst   & Cst   & Cst   & Htl   & Htl   & Htl   & Htl   & Cst   & Cst   & Cst   & Cst   & Cst   & Cst   & Cst   & Cst   & Cst   & Cst \\
    \textbf{CZE} & Cst   & Cst   & Cst   & Cst   & Met   & Met   & Met   & Met   & Met   & Met   & Met   & Met   & Met   & Met   & Obs   & Est   & Met \\
    \textbf{DEU} & Cst   & Cst   & Cst   & Cst   & Cst   & Cst   & Cst   & Cst   & Cst   & Cst   & Cst   & Cst   & Met   & Cst   & Cst   & Cst   & Met \\
    \textbf{DNK} & Cst   & Cst   & Cst   & Cst   & Obs   & Obs   & Obs   & Obs   & Cst   & Obs   & Obs   & Obs   & Obs   & Obs   & Obs   & Obs   & Obs \\
    \textbf{ESP} & Cst   & Cst   & Cst   & Cst   & Cst   & Cst   & Cst   & Cst   & Cst   & Cst   & Cst   & Cst   & Cst   & Cst   & Cst   & Cst   & Cst \\
    \textbf{EST} & Agr   & Agr   & Agr   & Agr   & Otr   & Elc   & Elc   & Elc   & Elc   & Elc   & Agr   & Cst   & Cst   & Obs   & Obs   & Obs   & Obs \\
    \textbf{FIN} & Cst   & Cst   & Cst   & Cst   & Cst   & Cst   & Cst   & Cst   & Cst   & Cst   & Cst   & Cst   & Cst   & Cst   & Cst   & Cst   & Cst \\
    \textbf{FRA} & Obs   & Obs   & Obs   & Obs   & Obs   & Obs   & Obs   & Obs   & Obs   & Obs   & Obs   & Obs   & Obs   & Obs   & Obs   & Obs   & Obs \\
    \textbf{GBR} & Fin   & Fin   & Fin   & Fin   & Fin   & Fin   & Fin   & Fin   & Fin   & Obs   & Obs   & Obs   & Obs   & Fin   & Obs   & Obs   & Obs \\
    \textbf{GRC} & Fod   & Fod   & Fod   & Fod   & Fod   & Cst   & Fod   & Cst   & Cst   & Cst   & Cst   & Cst   & Cst   & Cst   & Cst   & Cst   & Cst \\
    \textbf{HUN} & Agr   & Agr   & Agr   & Agr   & Elc   & Elc   & Fod   & Fod   & Fod   & Elc   & Elc   & Tpt   & Tpt   & Tpt   & Elc   & Elc   & Mch \\
    \textbf{IDN} & Agr   & Agr   & Agr   & Agr   & Agr   & Agr   & Agr   & Agr   & Agr   & Agr   & Agr   & Agr   & Agr   & Agr   & Agr   & Agr   & Agr \\
    \textbf{IND} & Agr   & Agr   & Agr   & Agr   & Agr   & Agr   & Agr   & Cst   & Cst   & Cst   & Cst   & Cst   & Cst   & Cst   & Cst   & Cst   & Cst \\
    \textbf{IRL} & Fod   & Fod   & Fod   & Fod   & Cst   & Cst   & Cst   & Cst   & Cst   & Cst   & Cst   & Cst   & Cst   & Cst   & Cst   & Chm   & Chm \\
    \textbf{ITA} & Met   & Met   & Met   & Met   & Met   & Cst   & Obs   & Obs   & Obs   & Obs   & Obs   & Obs   & Obs   & Obs   & Obs   & Obs   & Obs \\
    \textbf{JPN} & Cst   & Cst   & Cst   & Cst   & Cst   & Cst   & Cst   & Cst   & Cst   & Cst   & Cst   & Cst   & Cst   & Cst   & Obs   & Obs   & Obs \\
    \textbf{KOR} & Cst   & Cst   & Cst   & Cst   & Cst   & Cst   & Cst   & Cst   & Cst   & Cst   & Met   & Met   & Met   & Met   & Met   & Met   & Met \\
    \textbf{LTU} & Agr   & Agr   & Agr   & Agr   & Fod   & Agr   & Agr   & Ele   & Ele   & Ele   & Ele   & Ele   & Est   & Whl   & Ele   & Ele   & Ele \\
    \textbf{LUX} & Cst   & Cst   & Rub   & Cst   & Fin   & Fin   & Fin   & Fin   & Fin   & Fin   & Fin   & Fin   & Fin   & Fin   & Fin   & Fin   & Fin \\
    \textbf{LVA} & Fod   & Agr   & Agr   & Cst   & Cst   & Cst   & Cst   & Cst   & Cst   & Cst   & Cst   & Cst   & Cst   & Cst   & Cst   & Cst   & Cst \\
    \textbf{MEX} & Min   & Min   & Min   & Min   & Min   & Min   & Min   & Min   & Min   & Cok   & Cok   & Cok   & Cok   & Cok   & Cok   & Cok   & Min \\
    \textbf{MLT} & Elc   & Ele   & Elc   & Elc   & Elc   & Elc   & Ele   & Elc   & Elc   & Ele   & Ele   & Ele   & Ele   & Ele   & Ocm   & Ele   & Ele \\
    \textbf{NLD} & Obs   & Obs   & Obs   & Obs   & Obs   & Obs   & Obs   & Obs   & Cst   & Obs   & Obs   & Obs   & Obs   & Obs   & Cst   & Cst   & Est \\
    \textbf{POL} & Agr   & Agr   & Agr   & Agr   & Cst   & Fod   & Cst   & Cst   & Ele   & Cst   & Cst   & Cst   & Cst   & Cst   & Cst   & Cst   & Cst \\
    \textbf{PRT} & Cst   & Cst   & Cst   & Cst   & Cst   & Cst   & Cst   & Cst   & Cst   & Cst   & Cst   & Cst   & Cst   & Obs   & Obs   & Obs   & Obs \\
    \textbf{ROM} & Fod   & Fod   & Fod   & Fod   & Fod   & Fod   & Fod   & Fod   & Fod   & Fod   & Fod   & Fod   & Fod   & Fod   & Fod   & Fod   & Fod \\
    \textbf{RUS} & Ele   & Ele   & Ele   & Ele   & Ele   & Ele   & Ele   & Ele   & Ele   & Ele   & Cok   & Cok   & Cok   & Cok   & Cok   & Cok   & Cok \\
    \textbf{SVK} & Ele   & Ele   & Cst   & Ele   & Ele   & Cok   & Cok   & Tpt   & Tpt   & Tpt   & Tpt   & Tpt   & Tpt   & Ele   & Cst   & Cst   & Elc \\
    \textbf{SVN} & Cst   & Cst   & Cst   & Cst   & Cst   & Cst   & Cst   & Obs   & Cst   & Cst   & Cst   & Cst   & Cst   & Cst   & Cst   & Cst   & Cst \\
    \textbf{SWE} & Est   & Est   & Est   & Est   & Est   & Est   & Est   & Est   & Est   & Est   & Est   & Est   & Est   & Est   & Est   & Est   & Est \\
    \textbf{TUR} & Cst   & Cst   & Cst   & Fod   & Agr   & Agr   & Fod   & Agr   & Agr   & Agr   & Agr   & Fod   & Fod   & Fod   & Agr   & Agr   & Agr \\
    \textbf{TWN} & Cst   & Cst   & Cst   & Cst   & Cst   & Cst   & Cst   & Cst   & Cst   & Met   & Met   & Met   & Met   & Met   & Chm   & Met   & Met \\
    \textbf{USA} & Pub   & Pub   & Pub   & Obs   & Obs   & Obs   & Obs   & Obs   & Obs   & Obs   & Obs   & Obs   & Obs   & Obs   & Obs   & Obs   & Obs \\
    \botrule
    \end{tabular}%
    }
  \label{table_A8}%
\end{table}%

\end{document}